\documentclass[fleqn]{mnras}

\usepackage{graphicx}	
\usepackage{amsmath}	
\usepackage{amssymb}	
\usepackage{multicol}        
\usepackage{multirow}
\usepackage{bm}		
\usepackage{pdflscape}	
\usepackage{ulem}
\usepackage{caption}
\let\bibhang\relax              
\usepackage{natbib}
\usepackage{makecell}
\usepackage{array}
\usepackage{booktabs}

\newcommand{\msun}{\rm{M}_{\odot}}

\usepackage[T1]{fontenc}
\usepackage{ae,aecompl}

\usepackage{newtxtext,newtxmath}

\title[Machine-assisted Semi-Simulation Model: Estimating Baryonic Properties from Dark Matter]{Machine-assisted Semi-Simulation Model (MSSM):  \protect\\
Estimating Galactic Baryonic Properties from Their Dark Matter Using A Machine Trained on Hydrodynamic Simulations}

\author[Yongseok Jo \& Ji-hoon Kim]{
Yongseok Jo$^{1}$, 
Ji-hoon Kim$^{1
}$\thanks{Co: \href{mailto:me@jihoonkim.org}{me@jihoonkim.org}}
\\
$^{1}$Center for Theoretical Physics, Department of Physics and Astronomy, Seoul National University, Seoul 08826, Korea \\
}

\date{Last updated 2019 Feburary 1; in original form 2019 February 1}

\pubyear{2019}

\begin{document}
\label{firstpage}
\pagerange{\pageref{firstpage}--\pageref{lastpage}}
\maketitle

\begin{abstract}
We present a pipeline to estimate baryonic properties of a galaxy inside a dark matter (DM) halo in DM-only simulations using a machine trained on high-resolution hydrodynamic simulations. 
As an example, we use the {\sc IllustrisTNG} hydrodynamic simulation of a $(75 \,\,h^{-1}{\rm Mpc})^3$ volume to train our machine to predict e.g., stellar mass and star formation rate in a galaxy-sized halo based purely on its DM content. 
An extremely randomized tree (ERT) algorithm is used together with multiple novel improvements we introduce here such as a refined error function in machine training and two-stage learning. 
Aided by these improvements, our model demonstrates a significantly increased accuracy in predicting baryonic properties compared to prior attempts --- in other words, the machine better mimics {\sc IllustrisTNG}'s galaxy-halo correlation. 
By applying our machine to the {\sc MultiDark-Planck} DM-only simulation of a large  $(1 \,\,h^{-1}{\rm Gpc})^3$ volume, we then validate the pipeline that rapidly generates a galaxy catalogue from a DM halo catalogue using the correlations the machine found in {\sc IllustrisTNG}.  
We also compare our galaxy catalogue with the ones produced by popular semi-analytic models (SAMs). 
Our so-called machine-assisted semi-simulation model (MSSM) is shown to be largely compatible with SAMs, and may become a promising method to transplant the baryon physics of galaxy-scale hydrodynamic calculations onto a larger-volume DM-only run.   
We discuss the benefits that machine-based approaches like this entail, as well as suggestions to raise the scientific potential of such approaches. 
\end{abstract}




\begingroup
\let\clearpage\relax
\endgroup
\newpage
\section{INTRODUCTION}
\label{sec:intro}

Years of work have been devoted by numerous researchers to the gravitational $N$-body simulations which contains only dark matter (DM) in order to describe the evolution of large scale structures (LSS) in the Universe (e.g., Boylan-Kolchin et al. \citeyear{Boylan-Kolchin2009}; Klypin, Trujillo-Gomez \& Primack \citeyear{Klypin2011}; Angulo et al. \citeyear{Angulo2012}; Riebe et al. \citeyear{riebe2013multidark}; Watson et al. \citeyear{Watson2013}; Skillman et al. \citeyear{Skillman2014}; Heitmann et al. \citeyear{Heitmann2015}; Ishiyama et al. \citeyear{Ishiyama2015}). DM-only simulations also provide valuable insights into the spatial and velocity correlations (e.g., White et al. \citeyear{white1987clusters}, \citeyear{white1987galaxy}; Jenkins et al. \citeyear{jenkins1998evolution}), density profiles of individual halos (e.g., Navarro et al. \citeyear{navarro1997universal}; Bullock et al. \citeyear{bullock2001profiles}; Prada et al. \citeyear{prada2006far}, \citeyear{Prada2012}; Klypin et al. \citeyear{klypin2016multidark}),  angular momentum profiles and shapes (e.g., Cole et al. \citeyear{cole1996structure}; Lemson et al. \citeyear{lemson1999environmental}; Bullock et al. \citeyear{bullock2001universal}; Bett et al. \citeyear{bett2007spin}) and halo substructures (e.g., Moore et al. \citeyear{moore1999dark}; Klypin et al. \citeyear{klypin1999galaxies}; Springel et al. \citeyear{springel2008aquarius}; {Madau et al. \citeyear{madau2008dark}}).

However, gravitational dynamics alone is clearly not sufficient for understanding our Universe. 
Baryon physics must be taken into account via one of the two popular methods: hydrodynamic simulations, or semi-analytic models (SAMs). 
On the one hand, with the advent of high-performance computing units with a large amount of memories, fully hydrodynamics, high-resolution cosmological simulations have become one of the major tools in studying baryonic contributions in the Universe's evolution. 
Hydrodynamic simulations that treat baryon physics such as individual galaxy formation from $\sim$Mpc scales down to $\lesssim$100 pc scales have emerged in recent years despite the expensive computational costs. 
Prominent examples includes {\sc Illustris}  (Vogelsberger et al. \citeyear{vogelsberger2014illustris}, \citeyear{vogelsberger2014properties}; Genel et al. \citeyear{Genel2014-fd}), {\sc IllustrisTNG} (Pillepich et al. \citeyear{Pillepich2018-ou}; {Springel et al. \citeyear{Springel2018-hk}}; Nelson et al. \citeyear{Nelson2018-gv}), {\sc Horizon-AGN} (Dubois et al. \citeyear{dubois2014dancing}), {\sc Eagle} (Schaye et al. \citeyear{schaye2014eagle}), {\sc Romulus} (Tremmel et al. \citeyear{tremmel2017romulus}), {\sc Mufasa} (Dav\'e et al. \citeyear{dave2016mufasa}) and {\sc Simba} (Dav\'e et al. \citeyear{Dave2019-ts}).
On the other hand, in SAMs and empirical models, halos from DM-only simulations are ``colored'' with baryons based on relatively simple physical recipes (e.g., Baugh et al. \citeyear{Baugh2006-uc}; Benson \citeyear{Benson2010-wy}; Croton et al. \citeyear{croton2016SAGE}; Rodriguez-Puebla et al. \citeyear{rodriguez2017constraining}; Cora et al. \citeyear{cora2018SAG}; Moster et al. \citeyear{moster2018emerge}; Behroozi et al. \citeyear{behroozi2018universemachine}).  
While SAMs inevitably require a set of tunable parameters, the computational cost of typical SAMs is much less than that of high-resolution hydrodynamic simulations.  
In addition, SAMs make it easy to test and appreciate the importance of physical interactions and parameters in play (Silk \& Mamon \citeyear{Silk2012-yc}). 

Even with the cutting-edge computing technologies that have allowed us to simulate individual galaxies with high fidelity, the contemporary computation power is insufficient to describe a larger volume of the Universe (i.e., $\sim$Gpc scale) with detailed baryon physics resolved at $\lesssim$100 pc resolution. 
To obtain ``observable'' baryonic signatures populating such a large volume, combining DM-only simulations with a SAM has traditionally been the only strategy that is computationally feasible. 
But, now with the arrival of machine learning technology, preliminary studies have been carried out to combine DM-only simulations with machine learning algorithms such as random forest (RF) to produce galaxy catalogues (Kamdar et al. \citeyear{Kamdar2016-zm}; Agarwal et al. \citeyear{agarwal2018painting}; see also Kamdar et al. \citeyear{Kamdar2016-du}).   

Here, in what we call a machine-assisted semi-simulation model (MSSM), a machine --- suitable for big data regression --- is trained to first establish correlations between DM and baryonic properties in fully hydrodynamic simulations (e.g., DM mass and stellar mass in a halo). 
The machine is then tested and used to estimate various baryonic properties of a DM halo (either in hydrodynamic simulations or in DM-only simulations) based purely on its DM content.  
A well-constructed machine can generate an extensive galaxy catalogue out of a DM-only simulation of a large volume, within a fraction of time needed for a high-resolution hydrodynamic simulation.  
Furthermore, this method can be one of the most promising ways to accurately transplant the baryon physics of galaxy-scale hydrodynamic calculations (e.g., {\sc IllustrisTNG} in a $(75 \,\,h^{-1}{\rm Mpc})^3$ volume) onto a larger-volume DM-only simulation (e.g., {\sc MultiDark-Planck} in a $(1 \,\,h^{-1}{\rm Gpc})^3$ volume; {Klypin et al. \citeyear{klypin2016multidark}}).
Training the machine with a RF-type algorithm, we could also grasp the degree of contribution or ``feature importance'' by each of the input features (e.g., halo mass vs. halo angular momentum) in estimating a particular property (e.g., stellar mass). 
From the intuition gained by feature importances and by comparing the resulting catalogues with SAMs', we will be able to provide insights to improve the SAMs as well. 

In this article, we first focus on improving the machine training for MSSM, and compare our machine's accuracy with a simpler baseline model's (Sections \ref{sec:method} and \ref{sec:result-overview}). 
Major improvements include: a refined error function in machine training, using historical and environmental factors of a halo as inputs, and the two-stage learning with some predicted baryonic properties as an intermediary (Sections \ref{sec:preprocess} and \ref{sec:result-improve}). 
Among these, the logarithmic scaling in the error function alleviates the inaccuracy in the lower end of the predicted outputs. 
A scheme that ``links'' two machines is introduced; it uses a predicted output from one machine as an input to the next, and is found to be one of the most effective ways to enhance the MSSM's accuracy.  
Tested with the {\sc IllustrisTNG} dataset, our pipeline demonstrates a significantly increased accuracy in estimating baryonic properties than previous attempts do (Section \ref{sec:result1}). 
Our machine learning and application pipeline, MSSM, is shown to be largely compatible with popular SAMs when generating a galaxy catalogue using the DM-only simulation database {\sc MultiDark-Planck} (Section \ref{sec:result2}). 

The remainder of this paper is organized as follows. 
In Section \ref{sec:method}, we explain our methodology focusing on the pipeline of our machinery and the machine learning algorithm. 
The pre-processing scheme of input datasets is detailed, too.
In Section \ref{sec:result1}, we elaborate on how and how much our MSSM pipeline is improved when trained with the {\sc IllustrisTNG} dataset.   
Then in Section \ref{sec:result2}, we apply our machine to the  {\sc MultiDark-Planck} dataset, and compare our resulting galaxy catalogue with popular SAM catalogues.  
In Section \ref{sec:discuss}, we briefly point out a few technical issues of our model, and discuss how its scientific potential could be raised.
Finally we summarize and conclude the paper in Section \ref{sec:conclusion}.

\section{METHODOLOGY}
\label{sec:method}

\begin{figure*}
    \centering
    \vspace{-1mm}
    \includegraphics[width=0.93\linewidth]{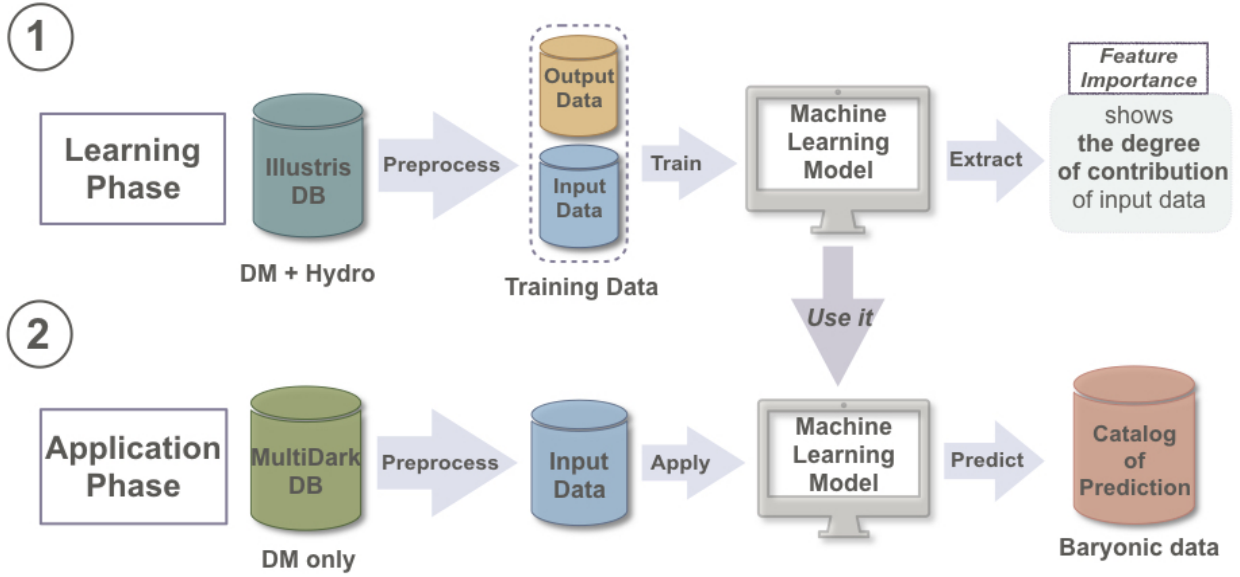}
    \vspace{-3mm}
    \caption[width=\paperwidth]{Flowchart of our machine-assisted semi-simulation model (MSSM).  
    In the learning phase ({\it top panel}), we train our machine with a fully hydrodynamic simulation database that contains both dark matter (DM) and baryon data (e.g., {\sc IllustrisTNG}) to predict the baryonic properties (``output'') based on the DM properties (``input''). 
    In the application phase ({\it bottom panel}), by feeding a DM-only $N$-body simulation (e.g., {\sc MultiDark-Planck}) to the trained machine, we produce a catalogue of baryonic predictions.
    See Section \ref{sec:flowchart} for more information about our MSSM pipeline.}
    \label{fig:structure}
    \vspace{0mm}
\end{figure*}

In this section, we describe the pipeline of our model and how we build and train our machine.   
In particular, we focus on the machine learning algorithm, and how we pre-process the input dataset to improve the machine's accuracy.

\subsection{Machine Learning Overview}
\label{sec:ML}

In our so-called MSSM, we exploit the results of fully hydrodynamic, high-resolution simulations to establish correlations or mappings --- not analytic prescriptions --- between DM and baryonic properties.  
Machine learning means training a machine for a task that typically deals with a large amount of data. 
If we assign two sets of data as ``input'' and ``output'', the machine by itself searches for a model and model parameters to take in the input and produce the output. 
In general, the more amount of data one gives, the more accurate the model becomes. 
The large datasets from modern cosmological simulations are thus ideal to exploit the novelty of machine learning.  

In the {\it supervised} learning phase of our work,\footnote{Machine learning algorithms are divided into several categories based on the amount and type of supervision in training: supervised learning, unsupervised learning, semi-supervised learning, and reinforcement learning.} we first divide the halo-galaxy catalogue from a large hydrodynamic simulation into a ``training set'' and a ``test set'' (see Section \ref{sec:phase1}). 
The machine learns a structure that maps an input to an output based on example input--output pairs, i.e., the training set (e.g., DM mass and stellar mass).
The machine looks for an optimized mapping by constantly evaluating the current mapping with an ``error function'' (or ``cost function''; e.g., a widely used metric in public packages is mean square error or MSE, see Section \ref{sec:ERT}). 
Based on this evaluation, the machine returns positive or negative feedback to itself. 
When the training is completed, one can ``score'' how well the machine can match the {\it actual} features in the simulation using the test set (see Section \ref{sec:phase2}). 
Based on this score, one may choose to update the learning algorithm or replace it with a different method. 

\subsection{Chosen Machine Learning Algorithm: Extremely Randomized Tree}
\label{sec:ERT}

The public machine learning package {\sc Scikit-Learn} offers an easy-to-use python interface and various hyper-parameters to adjust for a chosen regressor (Pedregosa et al. \citeyear{Pedregosa2011-iz}). 
We use the {\tt ExtraTreeRegressor} in {\sc Scikit-Learn}, an extremely randomized tree algorithm (ERT; {Geurts et al. \citeyear{geurts2006extremely}}).\footnote{ERT is chosen over other algorithms since the ``stacked'' multi-expert meta-regressor finds that ERT is almost always the most successful regressor.}
ERT is a branch of random forest (RF) algorithms which itself is a type of ensemble learning. 
We introduce the regressor's basic concept and inner workings here to later explain the improvements we made in the machine.

At the heart of an ERT lies a ``decision tree'' that is constructed top-down from a root node. 
The tree partitions the data into subsets which contain instances of similar values; a (leaf) node generally has more than one instance. 
A ``forest'' refers to an ensemble of decision trees --- i.e., a collection of trees makes a forest.  
Compared to a plain RF, ERT's additional randomization step arises as the tree nodes are split (i.e., the points of split are randomly chosen), which makes an ERT  perform mostly faster than a plain RF.  

To best split the nodes, different statistical techniques can be adopted, but a common choice is to use an error function (see Section \ref{sec:ML}). 
Often in the form of MSE, the error function helps determine the accuracy of an ERT model at each node as
\begin{equation}
    \text{MSE}_{\,\text{node}} = \frac{1}{N_{\text{node}}} \sum_{i \in\, {\text{node}}}{\left(y^{i} - y_{\text{node}} \right)^{2}},
    \label{eq:MSE_error}
\end{equation}
where $y_{\text{node}} = \frac{1}{N_{\text{node}}}\sum_{i \in \text{node}} y^{i}$, and $N_{\text{node}}$ is the number of instances at the node. 
It is important to employ an appropriate error function based on the data structure in use. 
MSE is the most common and widely used error function, but we note that in the reported study we choose a different metric to best serve our cosmological datasets. 
This will be discussed in detail in Section \ref{sec:result-logscale}.\footnote{In addition to the error function, other hyper-parameters in ERT include: maximum depth of a tree, minimum samples split, maximum number of nodes, etc.  The ``depth'' of a decision tree refers to the distance from a root node to a farthest leaf node.  The ``size'' of a tree is the number of all nodes. \label{foot:hyper}} 

One of the most salient advantages of ERT is that it is less prone to overfitting, a critical issue in machine learning. 
If we over-train the machine with a dataset of often a relatively small size, the machine could end up being skewed towards the particular input--output pairs. 
In other words, the machine may perform well on that particular datasets with high accuracy, but may not show similar accuracies when fed with different datasets. 
To mitigate overfitting, ERT uses subsets and boostrap aggregating (``bagging''; see {Geurts et al. \citeyear{geurts2006extremely}} for more information), and randomly splits nodes rather than looking for the least ``biased'' split points.\footnote{Therefore, when compared to RF, ERT decreases the ``variance'' of the model, but increase its ``bias''.  This is so-called bias--variance tradeoff.  High ``variance'' means that the machine is {\it overfitted} to random noises in a particular training set.  High ``bias'' means that the machine is {\it underfitted} that it only finds poor mappings between input and output data. \label{foot:variance}}
This way we could reduce the ``generalization error'' (as opposed to a ``sampling error'') when the machine is applied to previously unseen data.

\subsection{Flowchart of Machine-assisted Semi-Simulation Model (MSSM)}
\label{sec:flowchart}

The flowchart of our MSSM, the machine learning and application pipeline, is illustrated in Figure \ref{fig:structure}.
Our goal is to construct a machine to produce a galaxy catalogue by combining a DM-only $N$-body simulation and a machine learning technique, that is on a par with or better than catalogues made with popular SAMs. 
Our pipeline is divided into two main parts --- (1) the learning phase:  train a machine to estimate baryonic data out of DM data using a fully hydrodynamic simulation, and (2) the application phase:  apply the trained machine to a DM-only simulation to produce catalogues of galactic baryonic properties. 

\subsubsection{Learning Phase}
\label{sec:phase1}

In the learning phase, we use only the DM-related features extracted from the {\sc IllustrisTNG} hydrodynamic simulation of a $(75 \,\,h^{-1}{\rm Mpc})^3$ volume (``TNG100'' in Nelson et al. \citeyear{Nelson2018-gv}; see Section \ref{sec:TNG100} for more information) as input data.
We take these DM features such as DM halo mass and halo velocity dispersion as inputs, and the baryonic features such as stellar mass and gas mass of the halo as desired outputs.  
These input--output pairs --- a ``training set'' --- is used to train the machine via {\tt ExtraTreeRegressor} (Section \ref{sec:ERT}). 
Note that several historical and environmental characteristics of each halo not included in the native catalogue are computed in the pre-processing step (see Section \ref{sec:preprocess} and Table \ref{table:input} for more information). 
During the training process, 20\% of the {\sc IllustrisTNG} data is spared for a test --- a ``test set'' --- to score the accuracy of the machine afterwards. 
Fed with the test set, the resulting machine makes a set of predicted output data (e.g., stellar masses predicted  from  DM masses); and, by comparing it with the {\it actual} values in the simulation (e.g., the {\it actual} stellar masses in {\sc IllustrisTNG}) we ``score'' the machine. 
Common metrics for scoring the linear regression are MSE and Pearson correlation coefficient (PCC); but, in the reported study different measures are also used to evaluate the machine accuracy.  
We will discuss this in detail in Section \ref{sec:result-overview}.    

It is also worth to mention that ERT in our MSSM not only builds a map connecting inputs and outputs, but also provides the ``feature importance'' that shows which input feature contributes how much to predict a particular output (e.g., which input feature is more important to predict stellar mass, halo mass or halo angular momentum?).  
From the feature importance we may update the set of input parameters to increase the machine's accuracy, understand the underlying physics, and potentially provide insights to improve SAMs (see also Sections \ref{sec:intro} and \ref{sec:fimp}). 

\subsubsection{Application Phase}
\label{sec:phase2}

In the application phase, the machine from the learning phase is fed with a DM-only simulation data. 
Here, the {\sc MultiDark-Planck} DM-only simulation of a large  $(1 \,\,h^{-1}{\rm Gpc})^3$ volume is used as an input  (``MDPL2'' in {Knebe et al. \citeyear{knebe2018multidark}}; see Section \ref{sec:MDPL2} for more information). 
Needless to say, this input data needs to be pre-processed so that it is exactly in the same format and structure as the input used in the learning phase (Section \ref{sec:preprocess}).
A well-optimized machine can swiftly generate a galaxy catalogue once the DM-only simulation dataset is pre-processed.  
In our study, the machine is able to ``paint'' baryonic features on $\sim10^6$ halos in a large cosmological volume  in just a few tens of minutes. This is a miniscule amount of time when contrasted with what is typically needed for a high-resolution hydrodynamic simulation that resolves each galaxy-size halo with proper baryon physics.  
In Section \ref{sec:conclusion} we will discuss more on how to utilize MSSM for science.  

\newcolumntype{?}{!{\vrule width 1pt}}
\begin{table*}
\centering
\caption{All DM-related input parameters utilized to predict baryonic properties of a halo in our machine.  
See Section \ref{sec:preprocess} for more information.}
\label{table:input} 
\begin{tabular}{?cc||c|c|c?}
\Xhline{1pt}
&&Input Parameter &Definition&Graphical Description\\
\Xhline{1pt}
&\multicolumn{1}{|c||}{\multirow{3}{*}{\rotatebox[origin=c]{90}{Baseline}}}&
DM mass of a halo&Total mass of all DM particles bound to a halo& 
\multirow{4}{*}{\begin{minipage}{.28\textwidth}
    \vspace*{2mm}
    \centering
      \includegraphics[width=\linewidth, height=1cm, width=1.8cm]{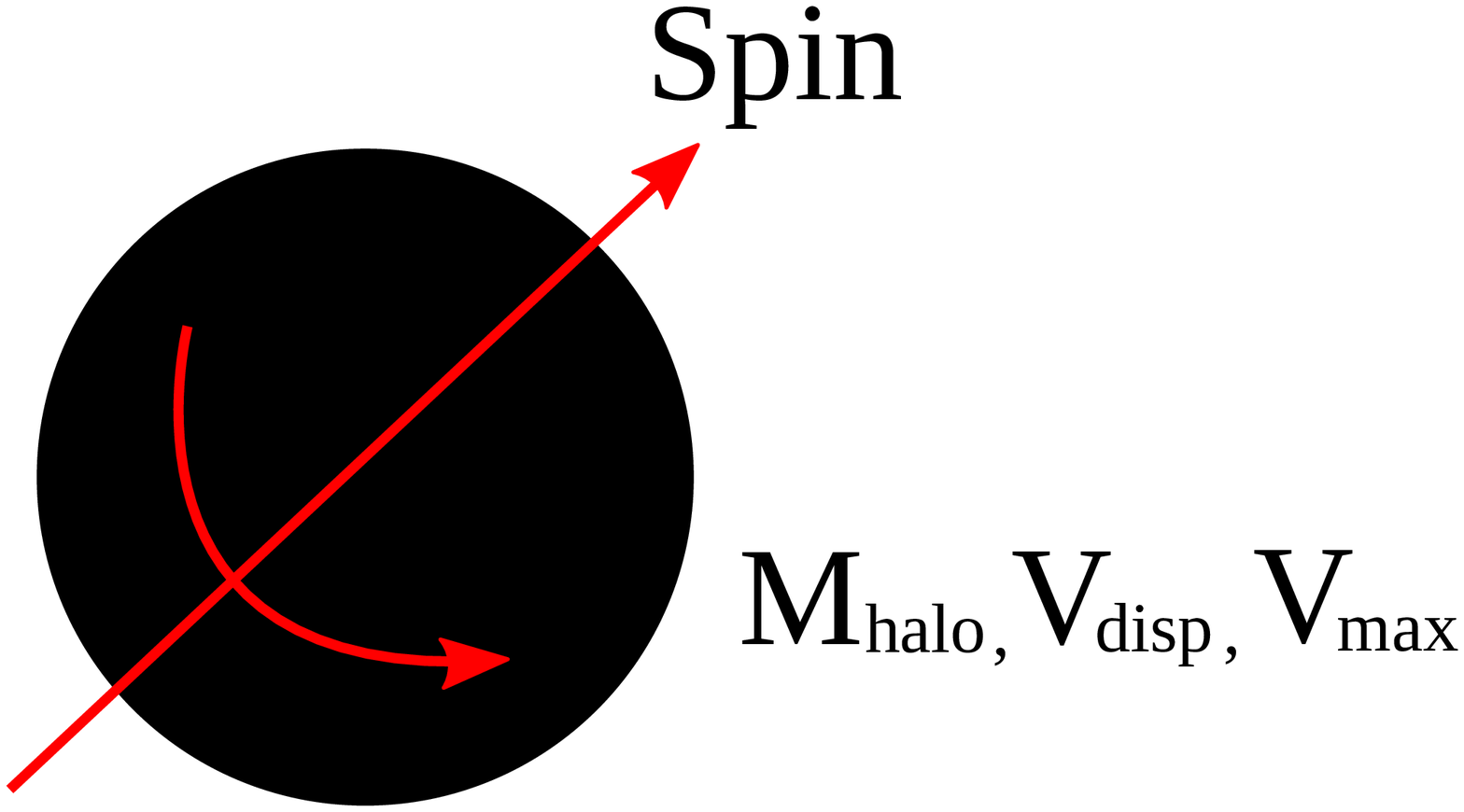}
    \vspace{2mm}
\end{minipage}}\\
\cline{3-4}
&\multicolumn{1}{|c||}{}&Velocity dispersion of a halo& Dispersion of all member particles' velocities&\\
\cline{3-4}
&\multicolumn{1}{|c||}{}&Maximum velocity of a halo& Maximum of spherically-averaged circular velocity&\\
\cline{2-4}
&&Angular momentum of a halo&Halo spin parameter&\\
\cline{3-5}
\multicolumn{1}{?r}{\multirow{8}{*}{}}

&&Number of all mergers&
\begin{minipage}{.3\textwidth}
    Number of all mergers throughout the halo's entire history
\end{minipage}&
\multicolumn{1}{c?}{
\begin{minipage}{.28\textwidth}
    \vspace*{2mm}
    \centering
      \includegraphics[width=\linewidth, height=1.6cm, width=3cm]{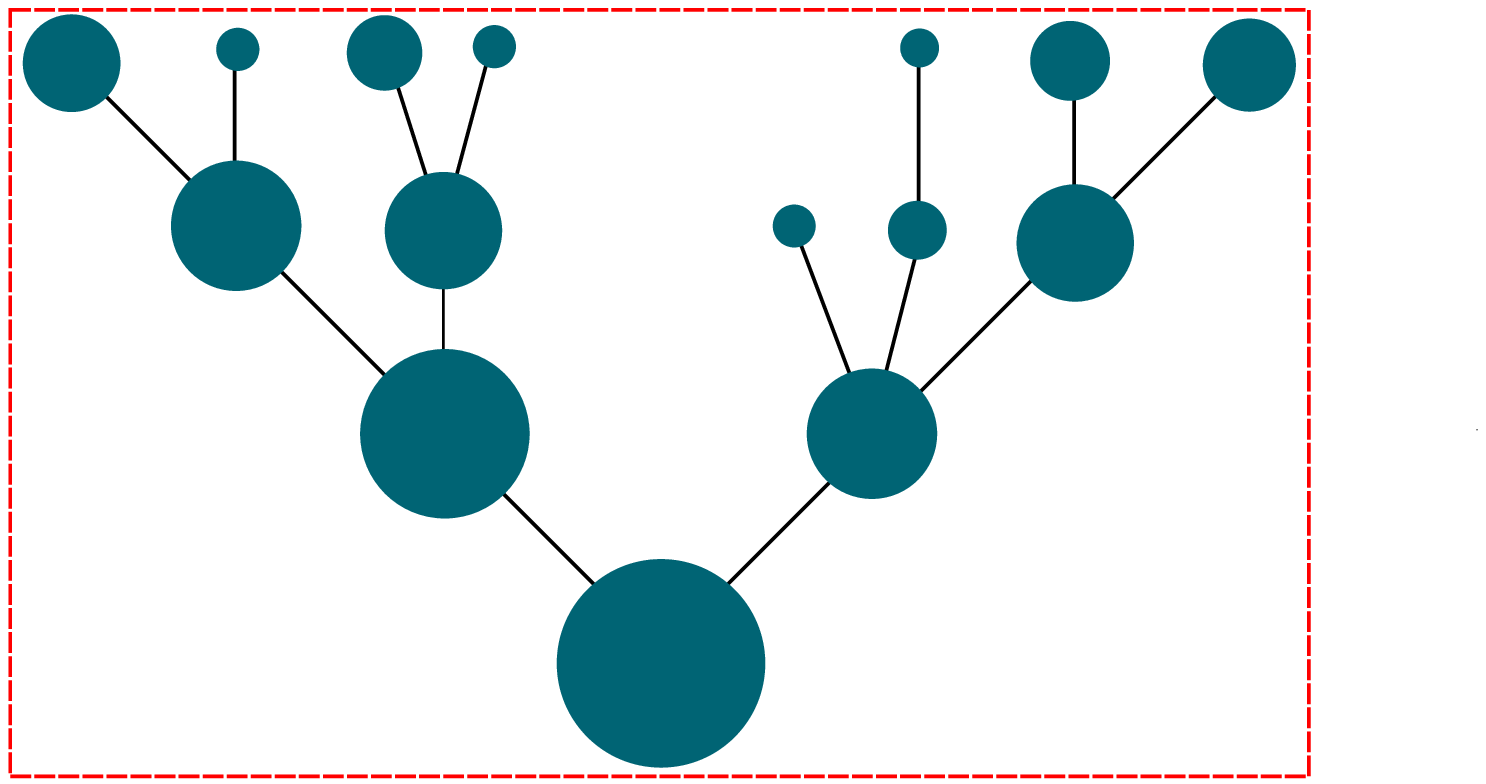}
    \vspace{2mm}
\end{minipage}}\\
\cline{3-5}

&&Number of all major mergers&
\begin{minipage}{.3\textwidth}
    Number of all mergers in which the mass ratios of the participating halos are less than 3:1
\end{minipage}&
\multicolumn{1}{c?}{
\begin{minipage}{.28\textwidth}
    \vspace*{2mm}
    \centering
      \includegraphics[width=\linewidth, height=1.6cm, width=3cm]{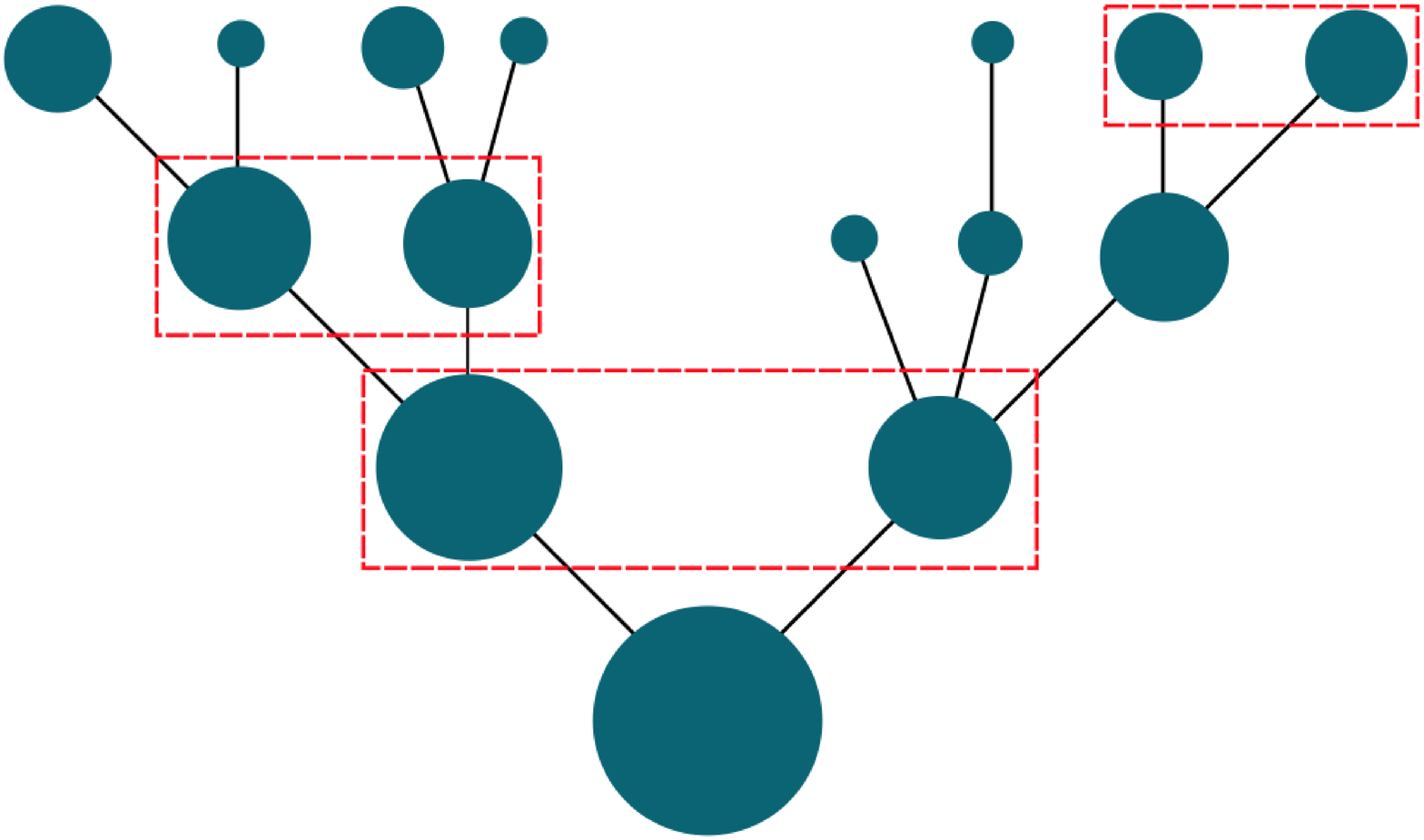}
    \vspace{2mm}
\end{minipage}}\\
\cline{3-5}

&&Last major merger mass ratio&
\begin{minipage}{.3\textwidth}
    The mass ratio of the most recent major merger along the merger tree
\end{minipage}&
\multicolumn{1}{c?}{
\begin{minipage}{.28\textwidth}
    \vspace*{2mm}
    \centering
    \includegraphics[width=\linewidth, height=1.6cm, width=3cm]{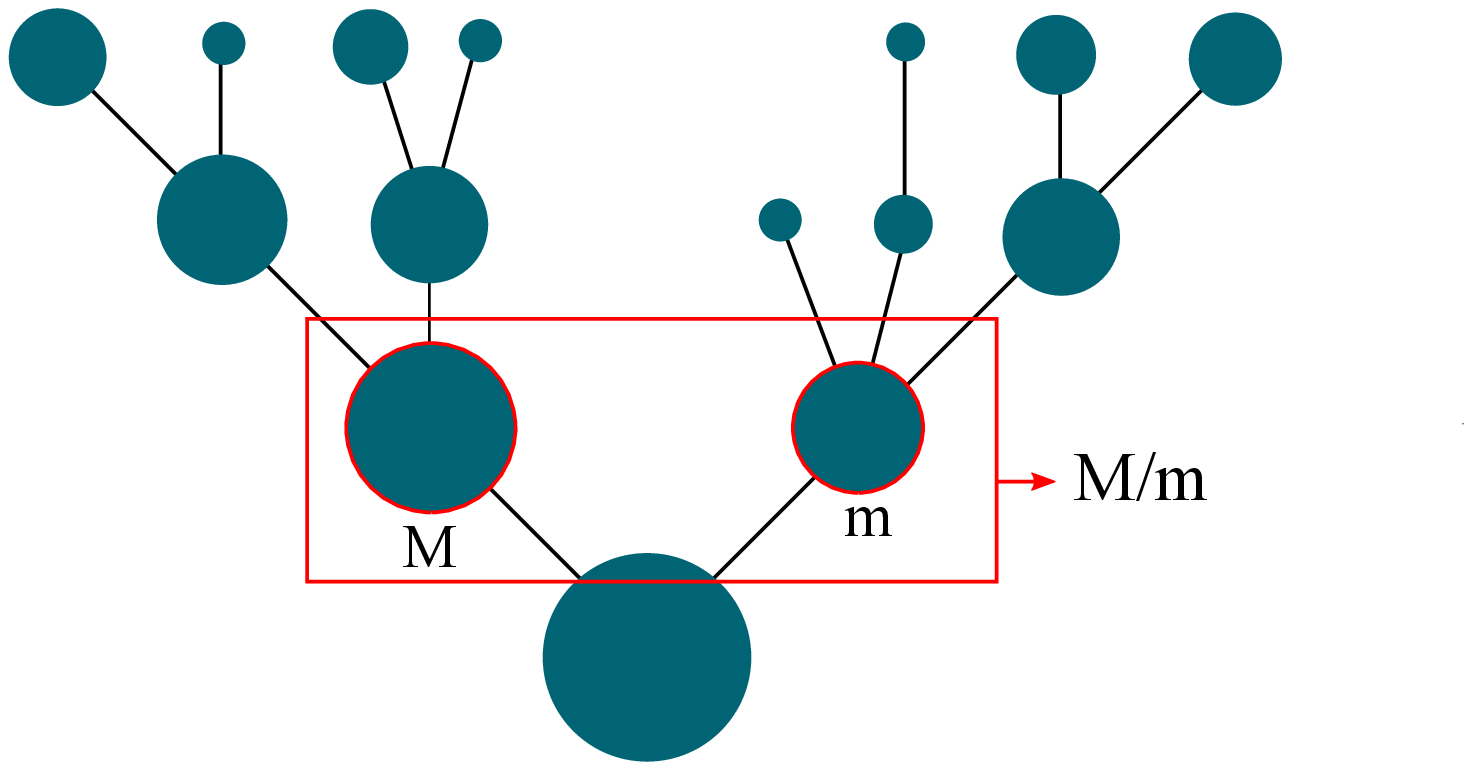}
    \vspace{2mm}
\end{minipage}}\\
\cline{3-5}

\rotatebox[origin=r]{90}{This Work}

&&\begin{minipage}{.1\textwidth}
    \centering
      \vspace*{6mm}
      Local density
\end{minipage}&
\begin{minipage}{.3\textwidth}
    \vspace*{6mm}
    The local density, $(\sum{M_{i}})/V_{\rm box}$, estimated for all  local halos within a $(2\, {\rm Mpc})^3$ volume
\end{minipage}&
\multicolumn{1}{c?}{
\multirow{2}{*}{
\begin{minipage}{.28\textwidth}
    \centering
      \vspace*{-2mm}
      \includegraphics[width=\linewidth, height=3cm, width=4.5cm]{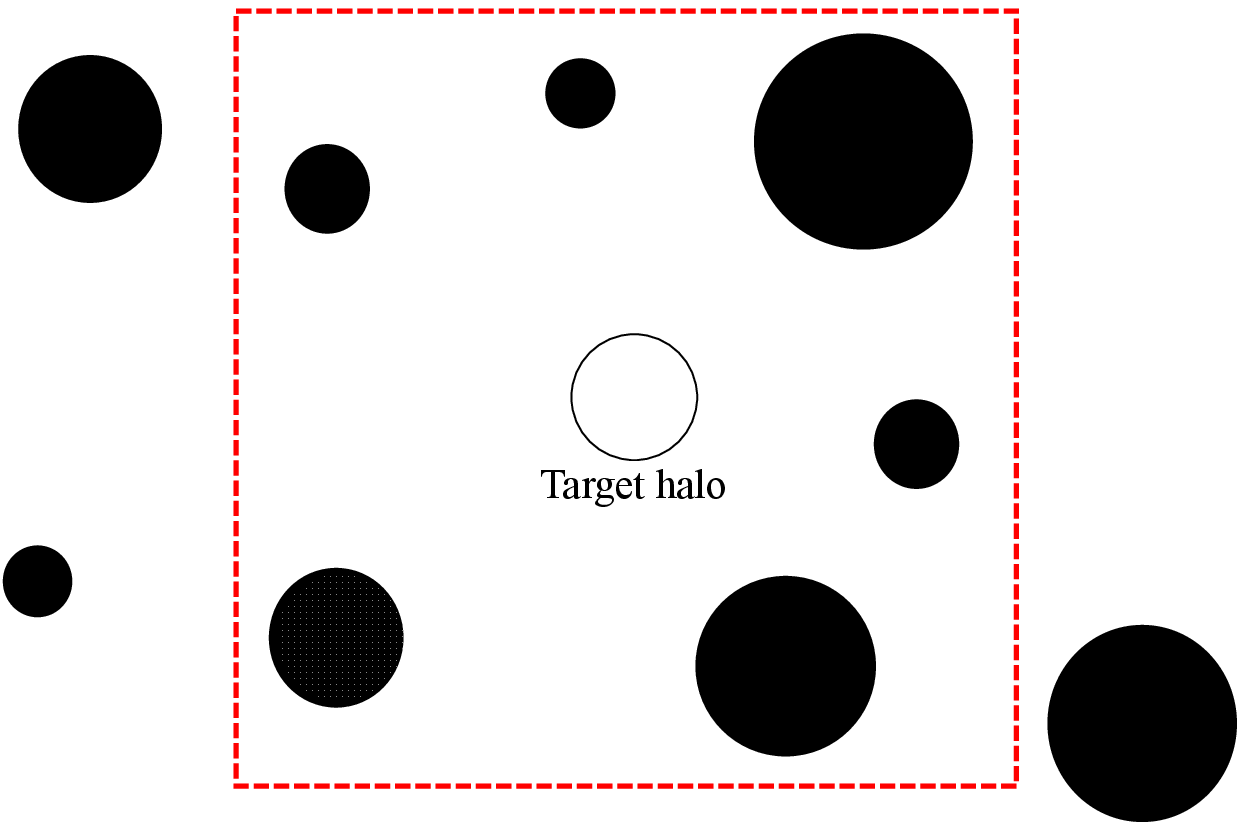}
      \vspace*{1mm}
\end{minipage}}}\\
\cline{3-4}
&&Number of local halos&
\begin{minipage}{.3\textwidth}
    \vspace*{5mm}
    Number of all local halos whose mass is larger than 80\% of the target halo's mass
    \vspace*{5mm}
\end{minipage}&\\
\cline{3-5}

&&Sum of mass over distance&
\begin{minipage}{.3\textwidth}
    Sum of mass over distance, $\sum M_{i} / R_{i}$, of all  local halos within a $(2\, {\rm Mpc})^3$ volume
\end{minipage}&
\begin{minipage}{.28\textwidth}
    \vspace*{2mm}
    \centering
      \includegraphics[width=\linewidth,height=3cm, width=4.5cm]{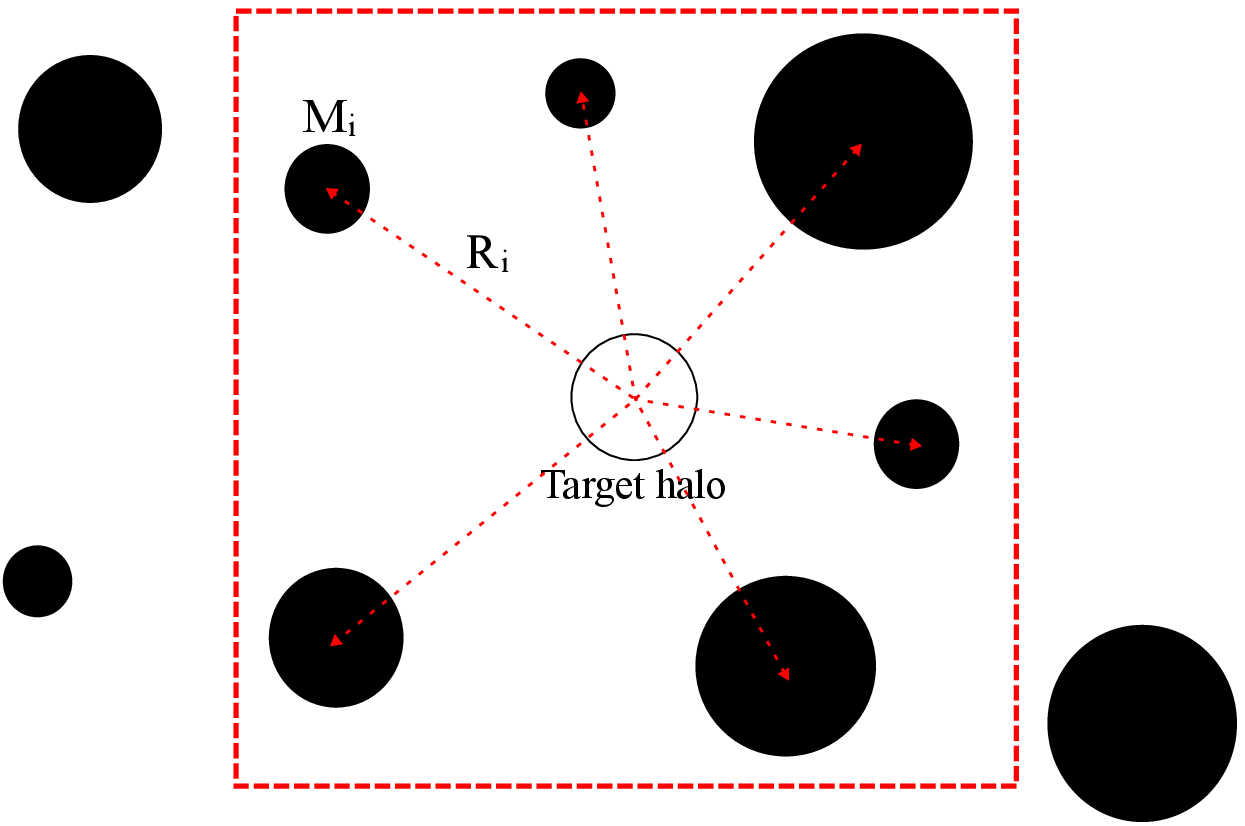}
    \vspace{2mm}
\end{minipage}\\
\cline{3-5}

&&Maximum mass over distance&
\begin{minipage}{.3\textwidth}
    Mass over distance, $M_{\rm max}/ R_{\rm max}$, for the most massive halo in the local volume
\end{minipage}&
\begin{minipage}{.28\textwidth}
    \vspace*{2mm}
    \centering
      \includegraphics[width=\linewidth,height=3cm, width=4.5cm]{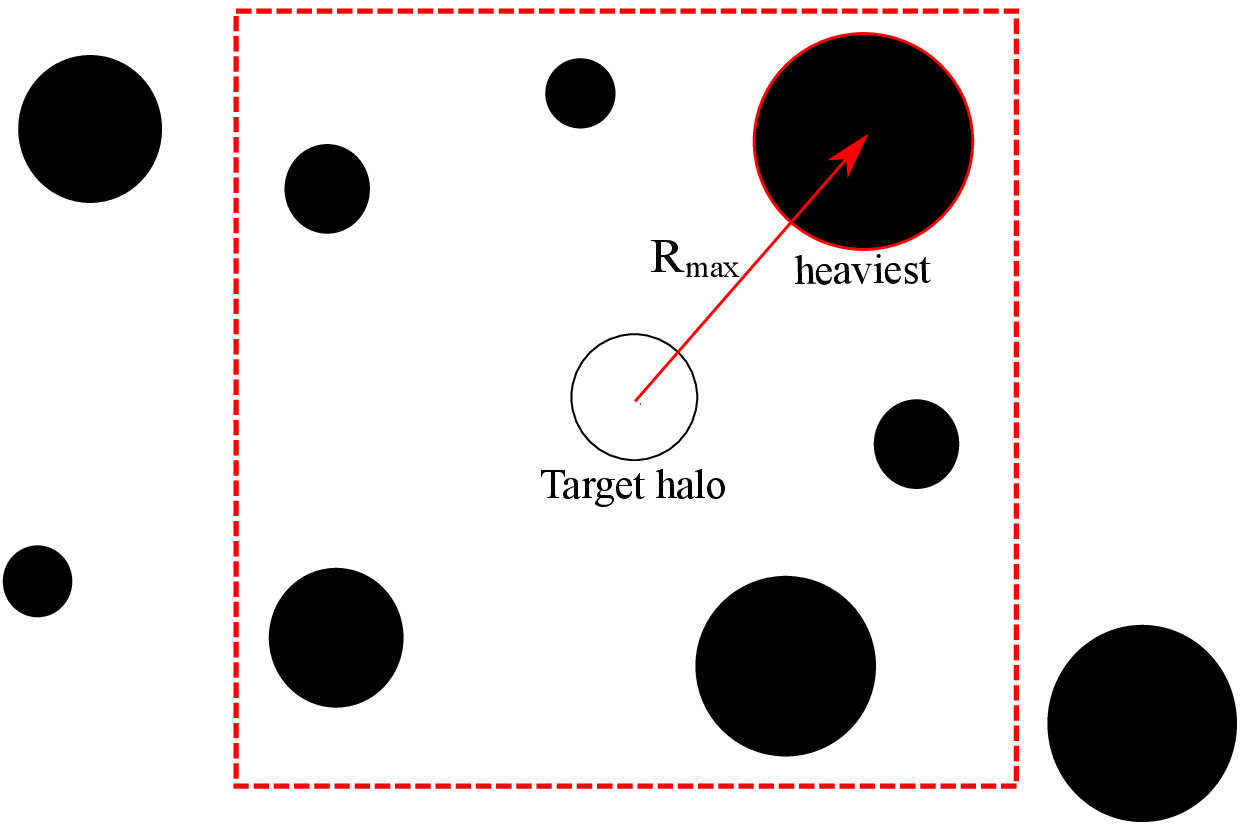}
    \vspace{2mm}
\end{minipage}\\
\Xhline{1pt}
\end{tabular}
\vspace{1mm}
\end{table*}

\subsection{Simulation Datasets for Machine Inputs}
\label{sec:data}

As noted in Section \ref{sec:flowchart} and Figure \ref{fig:structure}, two types of simulations are considered in our MSSM pipeline --- (1) in the learning phase: a fully hydrodynamic simulation is used to train a machine, and (2) in the application phase:  the trained machine is applied to a DM-only simulation to produce galaxy catalogues. 

\subsubsection{Hydrodynamic Simulation for The Learning Phase: {\sc IllustrisTNG}}
\label{sec:TNG100}

{\sc Illustris} (Vogelsberger et al. \citeyear{vogelsberger2014illustris}, \citeyear{vogelsberger2014properties}) and {\sc IllustrisTNG} (Pillepich et al. \citeyear{Pillepich2018-ou}; Nelson et al. \citeyear{Nelson2018-gv}) are gravito-hydrodynamic simulations performed with a moving-mesh code {\sc Arepo} (Springel \citeyear{Springel2010-jo}).
Both simulations include all relevant galaxy-scale physics to follow the evolution of dark matter, cosmic gas, stars and super massive black holes (SMBHs) from $z=127$ to 0, such as radiative gas cooling (Katz et al. \citeyear{Katz1995-hl}; Wiersma et al. \citeyear{Wiersma2009-kx}), star formation (Springel \& Hernquist \citeyear{Springel2003-tp}; Schaye \& Dalla Vecchia \citeyear{Schaye2008-on}), stellar evolution and chemical enrichment based on stellar synthesis models (Wiersma et al. \citeyear{Wiersma2009-ux}), stellar feedback (Springel \& Hernquist \citeyear{Springel2003-lt}) and SMBH and Active Galactic Nuclei (AGN) feedback  (Springel et al. \citeyear{Springel2005-br}, \citeyear{Springel2005-ip}). 
The more recent {\sc IllutrisTNG} (The Next Generation) updates {\sc Illustris} by including magneto-hydrodynamics (Pakmor et al. \citeyear{pakmor2011magnetohydrodynamics}; Pakmor \& Springel \citeyear{pakmor2013simulations}), metal advection (Naiman et al. \citeyear{naiman2018first}), updated SMBH physics (Wienberger et al. \citeyear{weinberger2017blackhole}; Weinberger et al. \citeyear{Weinberger2018-sw}), various computational improvements (detailed in Pillepich et al. \citeyear{Pillepich2018-ou}), as well as updated cosmology consistent with Planck Collaboration \citeyearpar{ade2016planck}: $\Omega_{\text{m},0}= 0.3089, \Omega_{\Lambda,0} =0.6911, \Omega_\text{b,0} = 0.0486, \sigma_8 = 0.8159, n_{\text{s}} = 0.9667$, and $h=$0.6774.

{\sc IllustrisTNG} is one of the most successful hydrodynamic calculations to date resolving individual galaxies with sophisticated baryon physics in a large enough volume. 
For this reason, we employ {\sc IllutrisTNG} in the learning phase of our MSSM pipeline (Section \ref{sec:phase1}).
In particular, among the three different box sizes the {\sc IllutrisTNG} database offers, the ``TNG100'' simulation of a $(75 \,\,h^{-1}{\rm Mpc})^3$ volume is adopted (``TNG100'' dataset as designated in Nelson et al. \citeyear{Nelson2018-gv}), where 100 denotes the simulation's approximate box size in Mpc). 
The TNG100 run was performed at three different resolutions: TNG100-1, -2 and -3 with TNG100-1 being the highest resolution run.  
At $z=127$, the TNG100-1 data consists of $1820^3$ DM particles with $m_{\text{DM}} = 7.5 \times 10^6 \,\text{M}_{\sun}$, and $1820^3$ hydrodynamic cells with ${m}_{\text{gas}}=1.4 \times 10^6 \,\text{M}_{\sun}$. 
At $z=0$ the simulation box holds 4371211 (sub)halos identified with the friends-of-friends halo finder (FOF; {Davis et al.\citeyear{Davis1985-ms}}) and the {\sc SubFind} subhalo finder ({Springel et al. \citeyear{Springel2001-vy}}). 
The publicly available halo catalogue also includes the merger trees built with the {\sc SubLink} code (Rodriguez-Gomez et al. \citeyear{Rodriguez-Gomez2015-pw}).\footnote{The {\sc IllustrisTNG} data is available at http://www.tng-project.org/.}  

\subsubsection{DM-only Simulation for The Application Phase: {\sc MultiDark-Planck}}
\label{sec:MDPL2}

{\sc MultiDark-Planck} (Riebe et al. \citeyear{riebe2013multidark}; Klypin et al. \citeyear{klypin2016multidark}; Rodr{\'\i}guez-Puebla et al. \citeyear{Rodriguez-Puebla2016}) is a DM-only gravitational dynamics simulation using {\sc L-Gadget-2}, a version of {\sc Gadget-2} optimized for a run with large number of particles ({Springel \citeyear{springel2005cosmological}}).
The cosmological model adopted is consistent with {Planck Collaboration (\citeyear{Ade2014-ve})}: $\Omega_{\text{m},0}= 0.3071, \Omega_{\Lambda,0} =0.6929, \Omega_\text{b,0} = 0.0482, \sigma_8 = 0.8228, n_{\text{s}} = 0.96$, and $h=$0.6777.

In the application phase of our MSSM  (Section \ref{sec:phase2}), the later version of {\sc MultiDark-Planck} is used as an input (``MDPL2'' dataset as designated in {Knebe et al. \citeyear{knebe2018multidark}}).
Run on a volume of $(1 \,\,h^{-1}{\rm Gpc})^3$ that is large enough to match observational surveys, MDPL2 depicts the large-scale evolution of a significant chunk of the Universe from $z=65$ to 0 using {$3840^3$} DM particles with $m_{\text{DM}} = 1.5 \times 10^9 h^{-1}\text{M}_{\sun}$ each. 
The MDPL2 database publicly provides a halo catalogue for each redshift snapshot identified with the {\sc Rockstar} code, along with the merger trees built with the {\sc Consistent Trees} code ({Behroozi et al. \citeyear{Behroozi2013-kk}}).\footnote{The {\sc MultiDark-Planck} data can be found in the {\sc CosmoSim} database at http://www.cosmosim.org/.}    


\subsection{Pre-processing The Simulation Datasets}
\label{sec:preprocess}

Data pre-processing is a pivotal step in machine learning. 
As noted in Figure \ref{fig:structure}, we transform the raw database --- the {\sc IllustrisTNG} data for the learning phase, and the {\sc MultiDark-Planck} data for the application phase --- into a desired input format for the machine.  

\subsubsection{Pruning The Input Datasets}
\label{sec:prune}
Becuase the resolutions of {\sc MultiDark-Planck} data and {\sc IllustrisTNG} data are different, to reconcile it we need to trim input datasets accordingly.  
MDPL2 dataset resolves dark matter halos down to $\sim 2.23 \times 10^{9} \,\msun$. 
TNG100-1 dataset resolves dark matter halos down to $7.5 \times 10^{\,6} \,\msun$ while resolving baryon down to $1.4 \times 10^{\, 6} \,\msun$. 
Therefore, we exclude the halos of  masses below $10^9 \,\msun$ in TNG100-1 to be compatible with MDPL2. 
In addition, since halos which do not contain star or gas are not our targets of interest, we have excluded halos whose stellar or gas mass is zero. 
With these cuts, the actual training set for the learning phase is reduced to $\sim 3\%$ of the original TNG100-1 halo catalogue. 
In Section \ref{sec:learning} we demonstrate that this training set is still sufficiently large for our learning process.

\subsubsection{Extracting Historical and Environmental Factors}
\label{sec:extract}

The ``baseline'' input features to predict baryonic properties include:  DM mass, velocity dispersion, and maximum circular velocity of a halo (see Table \ref{table:input}).  
This set of parameters --- straight from public halo catalogues --- is largely what prior attempts have used (e.g., Kamdar et al. \citeyear{Kamdar2016-zm}).  
In addition to the baseline parameters, in the present study we aim to capture what we refer to as ``historical'' and ``environmental'' factors, and add them to the input dataset.  
The new features for each halo are extracted (1) from the halo's merger history, and (2) from the halo's local volume.  

First, from the halo's merger tree, the following three features are obtained (Table \ref{table:input}):  the number of all mergers, the number of all major mergers, and the mass ratio of the last major merger.  
These characteristics are chosen to {\it explicitly} quantify the evolution history of a halo imprinted in the merger tree (unlike Agarwal et al. \citeyear{agarwal2018painting} where the merger-related parameters are implicit).  
Here, the mass ratio of participating halos must be less than 3:1 to be considered as major merger. 
Analogous to Rodriguez-Gomez et al. (\citeyear{Rodriguez-Gomez2015-pw}), the mass ratio is calculated when the secondary progenitor reaches its maximum halo mass, $t_{\rm max}$, before the two halos merge into one in the tree.  
We take this point $t_{\rm max}$ as the moment of merger.

Second, from the target halo's local volume of $(2\, {\rm Mpc})^3$, the following four features are extracted (Table \ref{table:input}):  the local density, the number of local halos whose masses are greater than 80\%  of the target halo's mass, the sum of mass over distance (``semi-potential'') of all local halos $\Phi_{\rm s} = \sum M_{i} / R_{i}$, and the mass over distance for the most massive local halo.
These parameters aim to characterize the target halo's local environment which has likely affected how the halo has evolved.  
Extracting these features from the raw dataset leads to the nearest neighbor search and range search problem. 
It requires us to construct a {\it k}-d tree that partitions the space into tree structure so that neighboring halos are efficiently located. 

Indeed, the {\it value-added} input datasets containing the additional input features improve the MSSM's accuracy for several feature predictions.  
This will be discussed in detail in Section \ref{sec:result-environment}. 

\begin{figure*}
    \includegraphics[width=0.91\textwidth]{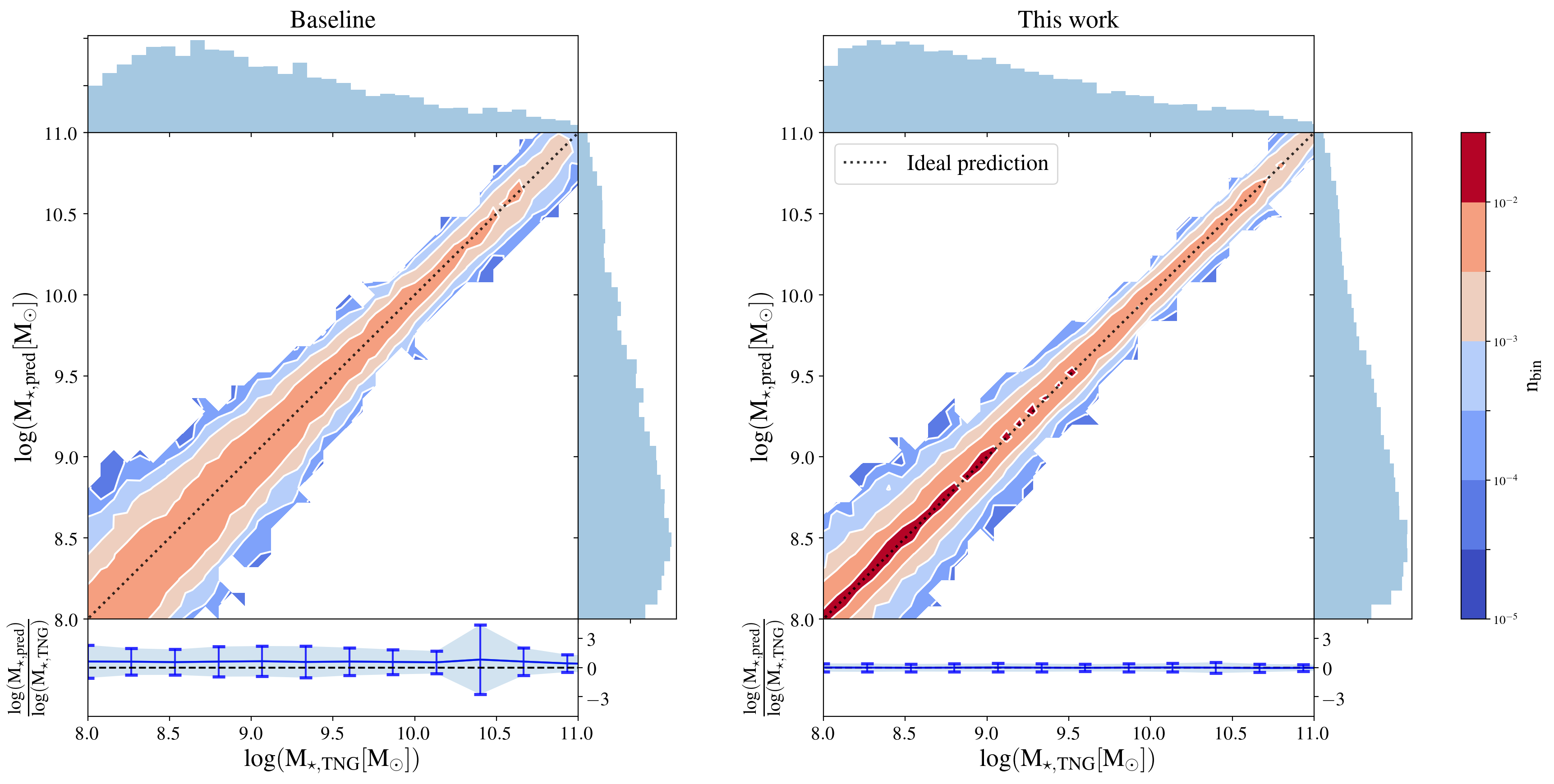}
    \vspace{1mm}
    \caption[width=\textwidth]{Normalized two-dimensional histogram comparing the {\it actual} stellar masses of halos in the {\sc IllustrisTNG} test set, $M_{\star, \text{TNG}}$, and the stellar masses predicted from input DM features of the test set,  $M_{\star, \text{pred}}$. 
    Colors indicate the normalized frequency, $n_{\text{bin}} = N_{\text{bin}}/N_{\text{tot}}$, where $N_{\text{tot}}$ is the total number of halos and $N_{\text{bin}}$ is the number of halos in each two-dimensional bin.
Results from two machine learning models are shown: the ``baseline'' model similar to previous studies ({\it left panel;} Section \ref{sec:extract}) and our model improved for its performance ({\it right panel;} see Sections \ref{sec:preprocess},  \ref{sec:result-improve} and Table \ref{table:input} for more information about their differences).  
    The black dotted line indicates an ideal prediction, $M_{\star, \text{pred}} = M_{\star, \text{TNG}}$.
    The marginal charts at the top and at the right show the distribution of $M_{\star, \text{TNG}}$ and $M_{\star, \text{pred}}$, respectively. 
    See Section \ref{sec:result-overview} for more discussion about this figure. 
    }
    \label{fig:comparison}
    \vspace{2mm}
\end{figure*}

\begin{figure*}
    \vspace{8mm}
    \includegraphics[width=0.92\textwidth]{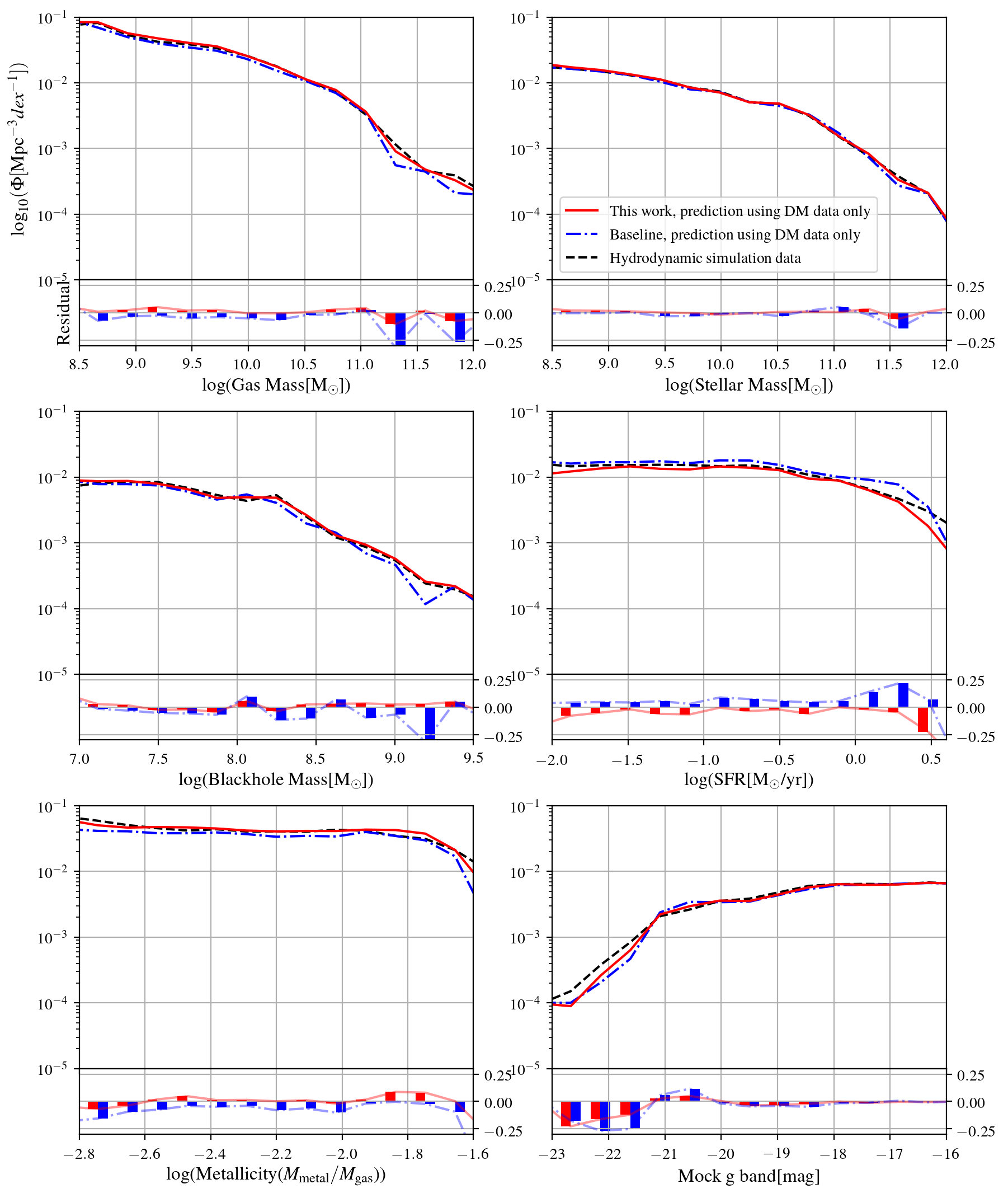}
    \vspace{4mm}
    \caption[width=\textwidth]{Probability distribution functions $\Phi$ (PDFs) of six baryonic properties --- gas mass, stellar mass, central black hole mass, star formation rate (SFR), metallicity, and stellar magnitude ($g$ band) --- predicted from input DM features in the {\sc IllustrisTNG} test set.  
    We use two machine learning models to make predictions: the baseline model ({\it blue dot dashed lines;} see Section \ref{sec:extract}) and our improved model ({\it red solid lines;} see Sections \ref{sec:preprocess},  \ref{sec:result-improve} and Table \ref{table:input} for more information about their differences). 
    The baryon data in {\sc IllustrisTNG} used for the training is also shown ({\it black dashed lines}).\textsuperscript{\ref{foot:how_to_pdf}}    
    The residuals between the predicted and the {\it actual} PDF, ${{\rm log}\,\Phi_{\,\text{pred}}}-{\rm log}\,\Phi_{\,\text{TNG}}$, are displayed in the bottom chart of each panel.  
    Overall, our model shows improved accuracy when predicting most baryonic properties of halos. 
    See Section \ref{sec:result1} for more discussion about this figure.
    }
    \label{fig:err_plot}
    \vspace{0mm}
\end{figure*}

\section{RESULTS 1: \,\, IMPROVING A MACHINE THAT PREDICTS BARYONIC PROPERTIES}
\label{sec:result1}

In Sections \ref{sec:result1} and \ref{sec:result2}, we present the results of our study focusing on the learning phase and the application phase of the MSSM pipeline (Figure \ref{fig:structure}), respectively.  
For the rest of the paper, unless the redshift of the data is specified, we only discuss the $z=0$ result.
We also note that we will focus on the halos of DM masses in the range of approximately $[10^{10}, 10^{13.5}] \,\msun$ when presenting our results in e.g., Figures \ref{fig:comparison} -- \ref{fig:ssfr_starmass} (but {\it not} necessarily when training the machine; see Section \ref{sec:prune}). 
It is because (1) for halos of DM masses below $10^{10} \,\msun$, the resolutions of {\sc IllustrisTNG} (Section \ref{sec:result1}) and {\sc MultiDark-Planck} simulations (Section \ref{sec:result2}) are too coarse for the machine to extract reliable mappings between DM and baryonic features, and (2) {\sc IllustrisTNG} contains insufficient number of halos of DM masses above $10^{13.5} \,\msun$ due to a small simulation box size.  
It should be noted that the limitation here is not about our model but about the available simulations; $[10^{10}, 10^{13.5}] \,\,\msun$ is indeed also the range for which the SAMs are best optimized.

\subsection{How Accurate Is The Machine's Prediction?} 
\label{sec:result-overview}

We first discuss how well our machine from the learning phase can predict halos' baryonic properties based purely on their DM features.  
Shown in Figure \ref{fig:comparison} are normalized two-dimensional histograms comparing the predicted stellar masses (``predicted output'') and the {\it actual} stellar masses  in a simulation (``desired output'' or ``answer''), when a test set from the {\sc IllustrisTNG} run is used.  
First, shown on left is the ``baseline'' model that considers only mass, velocity dispersion, and maximum circular velocity of a DM halo as inputs (similar to previous studies such as Kamdar et al. \citeyear{Kamdar2016-zm}; see Section \ref{sec:extract}). 
Shown on right is our model that improves the baseline one in various ways to be discussed in Section \ref{sec:result-improve}, including: a refined error function in machine training (Section \ref{sec:result-logscale}),  using historical and environmental factors of a halo as inputs (Sections \ref{sec:result-environment} and \ref{sec:extract}), and the two-stage learning with some predicted baryonic properties as an intermediary  (Section \ref{sec:result-photo}).  
We test both models to predict the following baryonic properties:  gas mass, stellar mass, central black hole mass, star formation rate (SFR), metallicity, and stellar magnitudes.\footnote{Stellar magnitudes are the luminosities of all star particles in eight photometric bands --- {\it U, B, V, K, g, r, i, z} --- as defined in Nelson et al. (\citeyear{Nelson2017-bz}).\label{foot:bands}}

Both histograms in Figure \ref{fig:comparison} are around the ideal prediction line ({\it black dotted line}), but in the bottom panels, the distribution is markedly tighter in our improved model resulting in the emergence of more concentrated region ({\it red region}) around the ideal prediction line. 
To quantify the machine's accuracy we first score each model with two common measures --- (1) mean square error (MSE),
\begin{equation}
   \text{MSE} = \frac{1}{N_{\rm tot}}\sum^{N_{\rm tot}}_{i} \left({y^{\,i}_{\text{pred}}}-{y}^{\,i}_{\text{TNG}}\right) \, , 
   \label{eq:MSE}
\end{equation} 
and (2) Pearson correlation coefficient (PCC), 
\begin{equation}
   \text{PCC} = \frac{\text{cov}\left(\,y_{\text{pred}}\,, \,y_{\text{TNG}}\right)}{\sigma_{y_{\text{pred}}} \sigma_{y_{\text{TNG}}}}\,,
   \label{eq:PCC}
\end{equation}
where $\text{cov}()$ is the covariance of two variables and $\sigma$ is the standard deviation. 
In both equations, $y^{\,i}_{\text{pred}}$ is the predicted {\it logged} output, and $y^{\,i}_{\text{TNG}}$ is the desired {\it logged} output in the simulation. 
Note that we take the logarithm of the output data because of the similar reason described in Section \ref{sec:result-logscale} --- except for stellar magnitudes where $y^{\,i}_{\text{pred}}$ and $y^{\,i}_{\text{TNG}}$ are simply the raw data (i.e., not logged).\footnote{Unlike other baryonic properties we consider, the stellar magnitudes are already logged and lie in the range of [-25, -13].  Therefore, the improvement for MSE or PCC suggested here in Section \ref{sec:result-overview}, or the proposed improvement in Section \ref{sec:result-logscale} is irrelevant for stellar magnitudes.\label{foot:bands-nolog}} 
We find that both measures are significantly improved in our model: MSE decreased from $2.0 \times 10^{-2}$ to $1.9 \times 10^{-4}$, and PCC increased from 0.971 to 0.987.  

We have also tried --- and eventually adopted --- another metric to measure the machine accuracy.\footnote{This is inspired by the case in which MSE or PCC does not aptly represent the entire $y^{\,i}_{\text{pred}} - y^{\,i}_{\text{TNG}}$ distribution --- i.e., PCC can be high even when the datapoints are widely spread out around the $y^{\,i}_{\text{pred}} = y^{\,i}_{\text{TNG}}$ line in Figure \ref{fig:comparison}.}
To compute what we call the ``mean binned error'' (MBE), first, the predicted and desired output pairs, $\left(y^{\,i}_{\text{pred}}, y^{\,i}_{\text{TNG}}\right)$, are binned into $\mathcal{N}_{\rm bins}$ bins according to the $y^{\,i}_{\text{TNG}}$ values. 
Then, in each bin, the {\it normalized} mean error is 
\begin{equation}
   \Gamma_{j} = \frac{1}{N_{j}}\sum^{N_{j}}_{i}\frac{\,\left|{y^{\,i}_{\text{pred}}}-{y}^{\,i}_{\text{TNG}}\,\right|\,}{{y^{\,i}_{\text{TNG}}}}\,, 
\end{equation} 
where $N_{j}$ is the number of data in the {\it j}-th bin.  
Finally, by averaging $\Gamma_{j}$'s across all bins we obtain the MBE as 
\begin{equation}
    \text{MBE} = \frac{1}{\mathcal{N}_{\rm bins}}\sum_{j}^{\mathcal{N}_{\rm bins}} \Gamma_j \,.
    \label{eq:MBE}
\end{equation}
If we replace the mean error in each bin, $\Gamma_{j}$, with the standard deviation in each bin, $s_{j}$, then we acquire another accuracy measure ``mean binned standard deviation'' (MBSD),
\begin{equation}
    \text{MBSD} = \frac{1}{\mathcal{N}_{\rm bins}}\sum_{j}^{\mathcal{N}_{\rm bins}} s_j \,.
     \label{eq:MBSD}
\end{equation}

We find that, in general, MBE captures the accuracy of a trained machine better than other metrics do.  
When predicting stellar masses, our model improves the MBE score from the baseline model's 0.0018 to 0.0013, and MBSD from 0.017 to 0.010.  
We will extensively use MBE and MBSD in Section \ref{sec:result-improve} and in Table \ref{table:score}.

In addition to reducing the machine accuracy down to a numeric score, we also inspect the machine's performance across the output's entire value range.  
In Figure \ref{fig:err_plot}, for six baryonic properties we predict, we compare the probability distribution functions (PDFs) of the two machine learning models, and the {\it actual} data in the simulation.\footnote{To make the PDF in Figure \ref{fig:err_plot}, we sum up the test results of 5 ($=1/0.2$) trials of machine learning and testing, where $0.2$ is the fractional size of the {\sc IllustrisTNG} test set (Section \ref{sec:phase1}).  Then, the fractional halo numbers in each bin match the number density in the real Universe.  For this reason, the black dashed line in Figure \ref{fig:err_plot} is slightly different from that of Figure \ref{fig:mdpl_pred}, the actual halo number density in the {\sc IllustrisTNG} volume (TNG100-1).  \label{foot:how_to_pdf}} 
Again, both the baseline ({\it blue dot dashed lines}) and our model ({\it red solid lines}) predict the baryonic properties well, but in general our improved model's PDFs better match the {\it actual} PDFs in {\sc IllustrisTNG} --- as demonstrated by the residual plots.  

\begin{table*}
\centering
\caption{The mean binned error (MBE), Eq. (\ref{eq:MBE}), quantifying how well the machine predicts each of the six baryonic properties --- gas mass, stellar mass, central black hole mass, SFR, metallicity, and stellar magnitude ($g$ band) --- based on DM features in the {\sc IllustrisTNG} test set. 
Each row indicates the MBE score within the respective $x$-range in Figure \ref{fig:err_plot} when the machine is improved by a single improvement --- except the ``Best combination'' row for which we identified the combination of improvements that yields the best scores for each prediction.\textsuperscript{\ref{foot:trials}} 
Numbers in the parentheses are mean binned standard deviation (MBSD), Eq. (\ref{eq:MBSD}).
See the referenced section in each row for details, and Section \ref{sec:result-improve} for more discussion about this table in general. 
}
\label{table:score} 
\vspace{2mm}
\begin{tabular}{lcccccc}
\hline
& Gas mass & Stellar mass & BH mass &  SFR & Metallicity & Stellar mag. ($g$)\\
\hline\hline
\multirow{2}{*}{Baseline (\S \ref{sec:extract}) }
&0.0015&0.0018&0.0047&1.71&0.022&0.0012\\
&(0.023)&(0.017)&(0.020)&(36.10)&(0.099)&(0.0121)\\
\hline
\multirow{2}{*}{Using an error function with logarithmic scaling (\S \ref{sec:result-logscale})}
&0.0010&0.0045&0.0126&1.70&0.010&\,\,--\textsuperscript{\ref{foot:bands-nolog}}\\
&(0.021)&(0.017)&(0.025)&(30.42)&(0.076)&(--)\\
\hline
\multirow{2}{*}{Using historical and environmental factors (\S \ref{sec:result-environment}, \S \ref{sec:extract})}
&0.0014&0.0014&0.0042&1.5&0.018&0.0010\\
&(0.023)&(0.016)&(0.018)&(28.27)&(0.093)&(0.0100)\\
\hline
\multirow{2}{*}{Two-stage learning (\S \ref{sec:result-photo})}
&0.0014&0.0016&0.0036&1.11&0.013&\,\,\,\,0.0005\textsuperscript{\ref{foot:mag-7bands}}\\
&(0.021)&(0.011)&(0.017)&(20.15)&(0.078)&(0.0064)\\
\hline
\multirow{2}{*}{Best combination (\S \ref{sec:result-combine})}
&0.0010&0.0013&0.0034&1.00&0.010&0.0005\\
&(0.020)&(0.010)&(0.016)&(20.23)&(0.070)&(0.0053)\\
\hline
\end{tabular}
\vspace{5mm}
\end{table*}

\subsection{Factors That Improved Our Model}
\label{sec:result-improve}

Having overviewed our machine's overall accuracy by comparing it with the {\it actual} data and with the baseline model, we now focus on each of the factors that improved our model.  
In the following sub-sections we explain each of three major improvements we made to our MSSM pipeline (Sections \ref{sec:result-logscale} - \ref{sec:result-photo}), followed by how we identify the best combination of these improvements that exhibits the highest accuracy (Section \ref{sec:result-combine}). 

\subsubsection{Using A Refined Error Function with Logarithmic Scaling}
\label{sec:result-logscale}

One of the most common choices for an error function in the machine learning algorithm --- including our choice, ERT --- is the MSE (see Section \ref{sec:ERT}), 
\begin{equation}
    \text{MSE}_{\,\text{node}} = \frac{1}{N_{\text{node}}} \sum_{i \in\, {\text{node}}}{\left(y^{i} - y_{\text{node}} \right)^{2}}.
    \tag{\ref{eq:MSE_error}}
\end{equation}
However, a severe problem may arise when our prediction target property has a large dynamic range (e.g., halo gas masses ranging from $10^{8}\,\text{M}_{\sun}$ to $10^{12}\,\text{M}_{\sun}$).
A simple mathematical argument tells that when naively used with raw $y$ values, MSE could be disproportionately more sensitive to larger $y$ values.
For example, a small fractional error in the $10^{12}\,\text{M}_{\sun}$ range may completely dominate over even a very large fractional error in the $10^{8}\,\text{M}_{\sun}$ range.
This has caused the naive baseline model (Section \ref{sec:extract}) to perform poorly in the lower value range (see e.g., the left panel of Figure \ref{fig:comparison}).
 
To amend the problem, in the learning phase, we apply {\it logarithmic scaling} to desired outputs  of the training set (i.e., actual baryonic properties in {\sc IllustrisTNG} --- except stellar magnitudes).\textsuperscript{\ref{foot:bands-nolog}}  
Or equivalently, the $y$ variables in the MSE error function, Eq. (\ref{eq:MSE_error}), now mean {\it logged} outputs, brining $y$ values to the range of $O(1)$. 
As a result, the equation is no longer heavily biased towards larger $y$ values.  
Hence, our fix alleviates the inaccuracy in the lower end of the predicted outputs (see e.g., the right panel of Figure \ref{fig:comparison}).\footnote{An alternative to the logarithmic scaling could be to normalize the raw $y$ values.  However, the normalized variables lose their physical meanings, so the physically meaningful quantities must be carefully recovered afterwards.  In contrast, logarithmic scaling does not lead to the loss of physical meaning.}
As seen in the 2nd row of Table \ref{table:score} where we assemble the scores by each of the improvements, predictions such as gas mass, SFR, and metallicity benefit from the refined error function (e.g., MBE for gas mass prediction decreased from 0.0015 to 0.0010).\footnote{Each of the MBE/MBSD scores in the table is an average over 200 trials.  A machine built in each trial is different due to the randomness in building an ERT, and in choosing a training set (80\% of the {\sc IllustrisTNG} data). \label{foot:trials}}
On the other hand, predictions for stellar and central black hole masses do not benefit as much from the refined error function alone.

\subsubsection{Using Historical and Environmental Factors}
\label{sec:result-environment}

As discussed in Section \ref{sec:extract}, we extract and add ``historical'' and ``environmental'' factors to the input features when we pre-process the data for our MSSM pipeline.   
The newly added features are extracted (1) from the halo's merger history, and (2) from the halo's local volume, aimed at directly and indirectly capturing the halo's evolution history.  
The resulting {\it value-added} dataset includes seven additional input features such as: number of all mergers, number of all major mergers, mass ratio of the last major merger, local density, number of local halos whose masses are greater than 80\%  of the target halo's mass, etc. (see Section \ref{sec:extract} for details).
It improves our model's MBE and MBSD scores when predicting features like stellar mass, central black hole mass, and SFR (see the 3rd row of Table \ref{table:score}).
For other features, including these extra factors is not as effective by itself. 

\subsubsection{Two-stage Learning With Stellar Magnitudes As An Intermediary}
\label{sec:result-photo}

Broadly speaking, the accuracy of the ERT machine learning algorithm improves as the number of decision trees or the ``size'' of each tree increases (Section \ref{sec:ERT}).\textsuperscript{\ref{foot:hyper}}
Since the increased tree size  requires exponentially more computing resources, we often need to limit the ``depth'' of a tree, and/or prune the nodes that are not functional.  
In practice, however, it is difficult to grow a large tree and prune them into an efficient shape. 

Here we introduce a scheme that ``links'' two machines, by using a predicted output from one machine as an input to the next.  
The ``two-stage learning'' scheme works as follows. 
Imagine building a machine to predict SFR based only on DM features (e.g., DM mass or velocity dispersion).  
To increase the machine accuracy the tree must be both deep and large, requiring copious computing resources.  
A training set may not be informative enough for a machine to establish a meaningful direct mapping between the DM properties and SFR within a practical time limit.  
Instead, here we first build a machine estimating stellar magnitudes based on DM properties, then use the predicted stellar magnitudes as part of inputs to another machine estimating SFR.\textsuperscript{\ref{foot:bands}}$^{,}$\footnote{To predict the $g$ band, the other seven bands are used to link the machines.\label{foot:mag-7bands}} 
By supervising what to estimate first (stellar magnitudes) in order to predict the eventual output (SFR), we effectively ``guide'' the machine to build one combined, large --- yet efficient --- ERT.  
Readers should note that we select stellar magnitudes as an ``intermediary'' because (1) the stellar magnitudes are relatively accurately predicted only from DM features and (2) the stellar magnitudes and SFR are highly correlated in the simulation data.\footnote{For example, SFR is more strongly correlated with the stellar magnitudes (e.g., $g$ band) than with any other DM features like DM mass.  In other words, when predicting SFR, stellar magnitudes' ``feature importances'' dominate ($> 50\%$; see Section \ref{sec:fimp}) over other DM features'.}
Thus, we argue that in the two-stage machine training, new astrophysical information is provided to the machine by a human supervisor that the stellar magnitudes are a good intermediary between DM properties and SFR. 
For more discussion on how stellar magnitudes and the two-stage learning can improve the performance of MSSM, see Appendix \ref{app:mag}.

We find that the two-stage learning technique described here is one of the best strategies to construct a large and efficient ERT, and is also arguably the most effective way to improve the MSSM's accuracy. 
As an example, for the SFR prediction,  the two-stage learning scheme improves both MBE and MBSD scores the most when compared with any other improvements (e.g., MBE for SFR prediction decreased from 1.71 to 1.11, and MBSD from 36.10 to 20.15; see the {\it 4th row} of Table \ref{table:score}).

\begin{figure*}
    \vspace{8mm}
    \includegraphics[width=0.92\textwidth]{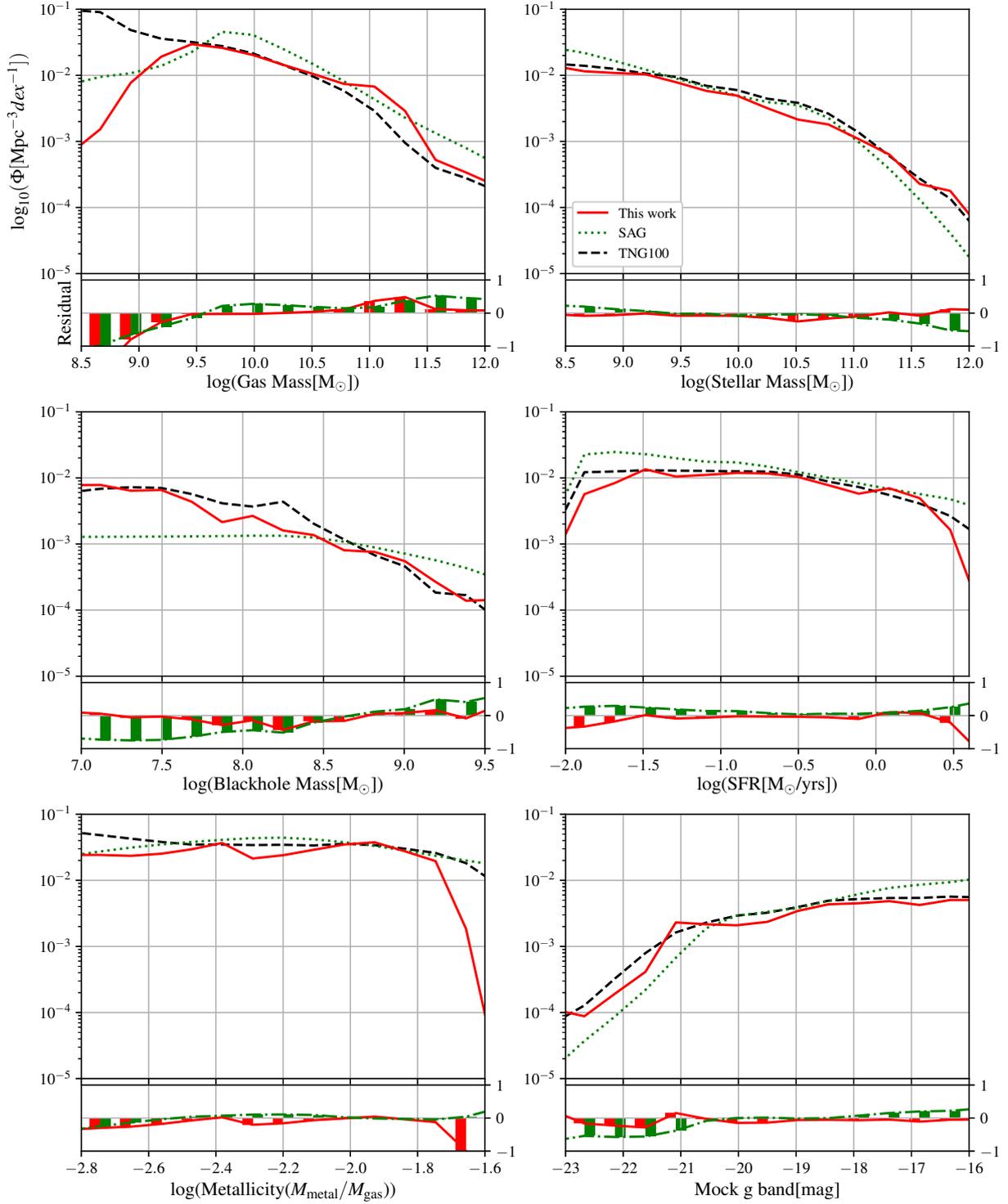}
    \vspace{4mm}
    \caption[width=\textwidth]{Probability distribution functions $\Phi$ (PDFs) of six baryonic properties predicted using a DM halo catalogue from the {\sc MultiDark-Planck} database. 
    Our improved machine trained with {\sc IllustrisTNG} is applied to a {\sc MultiDark-Planck} dataset to make predictions ({\it red solid lines;} see Figure \ref{fig:structure} and Sections \ref{sec:preprocess}, \ref{sec:result-improve} about our improved model).
    We compare our prediction with a catalogue by a semi-analytic model (SAM) code {\sc Sag} ({\it green dotted lines;} see Section \ref{sec:result2}). 
    The {\it actual} baryon data in the {\sc IllustrisTNG} itself is also shown (TNG100-1; {\it black dashed lines}).\textsuperscript{\ref{foot:how_to_pdf}}  
    The residuals between the predicted PDF and the simulation's PDF ({\sc IllustrisTNG}'s), ${{\rm log}\,\Phi_{\,\text{pred}}}-{\rm log}\,\Phi_{\,\text{TNG}}$, are displayed in the bottom chart of each panel.  
    Our machine-assisted semi-simulation model (MSSM) and the SAM show compatible results overall when assigning baryonic properties to halos.  
    See Section \ref{sec:result-compatible} for more discussion about this figure.
    }
    \label{fig:mdpl_pred}
    \vspace{0mm}
\end{figure*}
 
\begin{figure*}
    \includegraphics[width=0.83\textwidth]{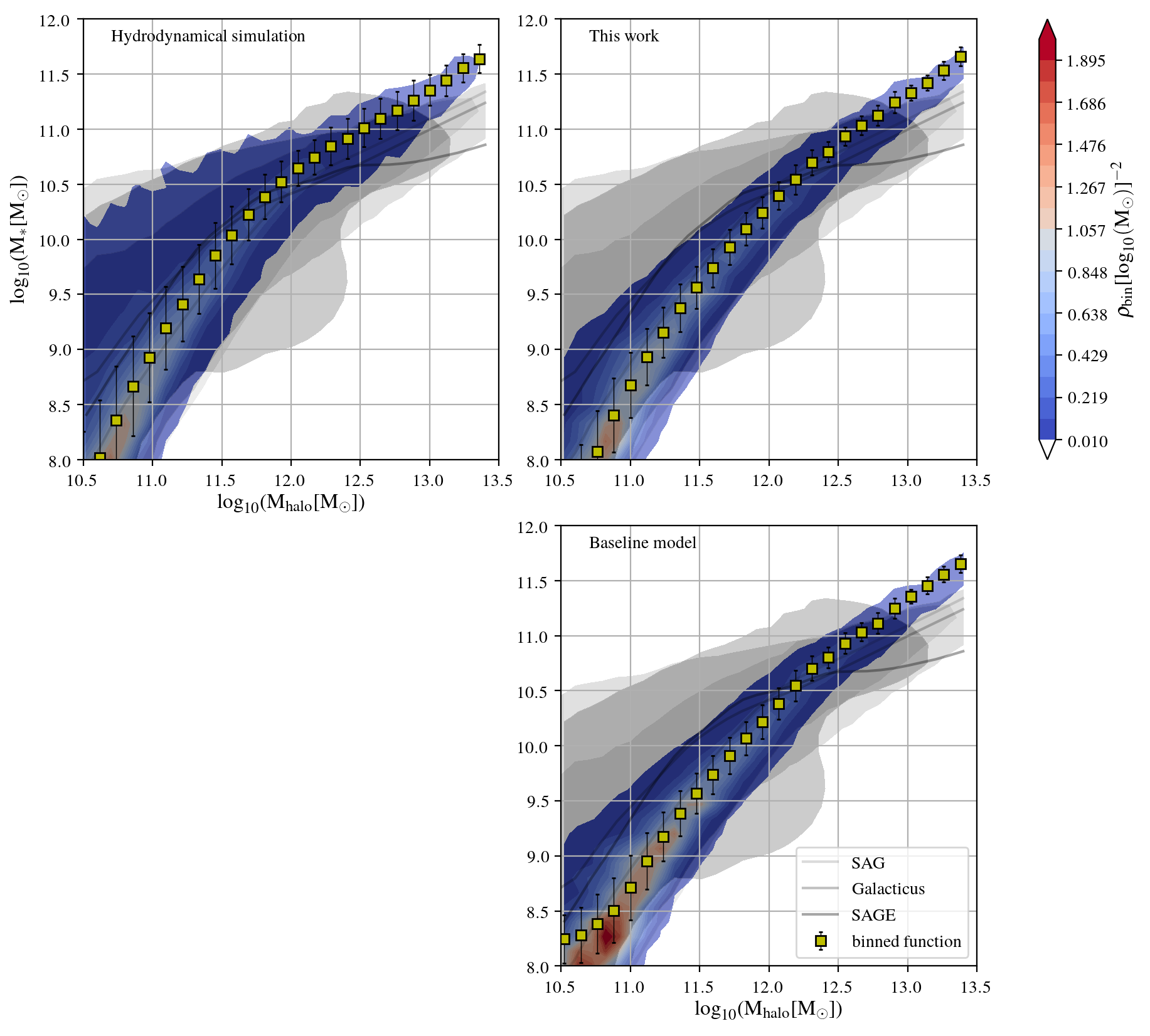}
    \vspace{0mm}
    \caption[width=\textwidth]{Two-dimensional probability distribution of DM halo masses, $M_{\rm halo}$, and predicted stellar masses,  $M_{\star}$ at $z=0$. Colors indicate $\rho_{\text{bin}} = N_{\text{bin}}/(N_{\text{tot}}S_{\text{bin}})$, where $N_{\text{tot}}$ is the total number of halos, $N_{\text{bin}}$ is the number of halos in each two-dimensional bin, and  $S_{\text{bin}}$ is the bin area.  
    Machines trained with {\sc IllustrisTNG} are applied to a {\sc MultiDark-Planck} dataset to make the PDF predictions:  the baseline model ({\it bottom right panel}) and our improved model ({\it top right panel;} see Sections \ref{sec:preprocess},  \ref{sec:result-improve} and Table \ref{table:input} for their differences).  
    Yellow squares represent binned averages.  
    The {\it actual} baryon data in the {\sc IllustrisTNG} itself is also presented ({\it top left panel}).  
    Shown in each panel as gray contours are results by three popular SAMs:  {\sc Sag}, {\sc Sage}, and {\sc Galacticus}. 
    See Section \ref{sec:result-compatible} for more discussion about this figure. 
    }
    \label{fig:mstarmhalo}
    \vspace{0mm}
\end{figure*}

\begin{figure*}
    \includegraphics[width=0.85\textwidth]{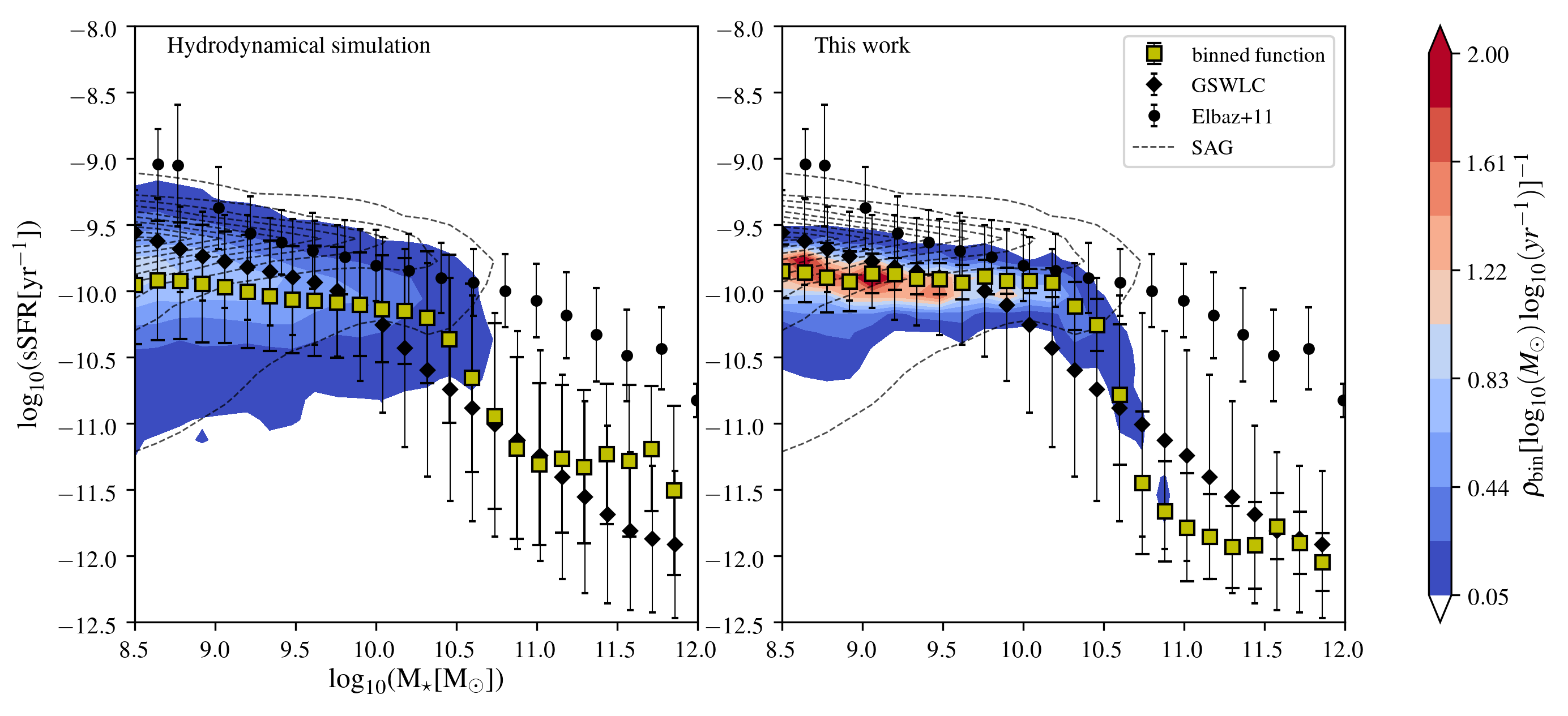}
    \vspace{0mm}    
    \caption[width=\textwidth]{Two-dimensional probability distribution of predicted stellar masses, $M_{\star}$, and  predicted specific SFRs at $z=0$. 
    Colors indicate $\rho_{\text{bin}} = N_{\text{bin}}/(N_{\text{tot}}S_{\text{bin}})$, where $N_{\text{tot}}$ is the total number of halos, $N_{\text{bin}}$ is the number of halos in each two-dimensional bin, and  $S_{\text{bin}}$ is the bin area. 
    Yellow squares represent binned averages, black diamonds represent {\it GALEX}-SDSS-{\it WISE} Legacy Catalog (GSWLC) from Salim et al. \citeyear{salim2016} at $z\sim0$, and black circles represent a compilation of observations from Elbaz et al. \citeyear{Elbaz2011-eu} at $z\sim0$.
    Our machine trained with {\sc IllustrisTNG} is applied to a {\sc MultiDark-Planck} dataset to  predict the PDF ({\it right panel}).  
    The {\it actual} baryon data in the {\sc IllustrisTNG} itself is also presented ({\it left panel}).  
    Shown in each panel as black dotted contours is the result by a SAM code, {\sc Sag}.
    See Section \ref{sec:result-underfit} for more discussion about this figure.      
    }
    \label{fig:ssfr_starmass}
    \vspace{0mm}
\end{figure*}

\subsubsection{Combining Improvements To Construct The Best Model}
\label{sec:result-combine}

Finally, we combine all three improvements discussed above. 
Rather than using all the improvements at once, we have carefully tested various combinations of improvements per each of baryonic properties.  
This is because, when combined, one improvement may hurt the other and lead to an unexpected decrease in machine accuracy.
The MBE scores for the identified best combinations are shown in the last row of Table \ref{table:score}.
The best combinations identified here have been referred as our ``improved model'', and are used to produce Figures \ref{fig:comparison} -- \ref{fig:ssfr_starmass}. 

In Table \ref{table:score}, readers may notice that the score of a best combination is sometimes the same as that of a single improvement.  
For example, the MBE for a stellar magnitude prediction is 0.0005 for the best combination, but also for the two-stage learning alone.  
This means that the two-stage learning technique is the most important and dominant factor in improving the accuracy of stellar magnitude prediction.  

\vspace{10mm}
\section{RESULTS 2: \,\,\,  PREDICTING BARYONIC PROPERTIES IN DARK MATTER-ONLY SIMULATIONS}
\label{sec:result2}
\vspace{2mm}

We now turn to the application phase of our MSSM pipeline (Figure \ref{fig:structure}), and use the machine to estimate baryonic properties for halos in a DM-only $N$-body simulation data. 
The machine from Section \ref{sec:result1} trained with the {\sc IllustrisTNG} data in the learning phase, is fed with the {\sc MultiDark-Planck} DM-only simulation (MDPL2; see Section \ref{sec:MDPL2}).\footnote{We note that the DM halos in DM-only simulations and hydrodynamic simulations have experienced different physical processes so are inevitably different.  But we also note that the so-called baryonic back-reaction effect is relatively small, justifying our use of a machine trained with hydrodynamic simulations in a different domain.  For more discussion, see Appendix \ref{app:baryon}.} 
The machine is asked to generate a galaxy catalogue with multiple baryonic properties --- gas mass, stellar mass, central black hole mass, SFR, metallicity, and stellar magnitudes --- filling the entire {\sc MultiDark-Planck}  volume of $(1 \,\,h^{-1}{\rm Gpc})^3$\footnote{The halo catalog of our Machine-assisted Semi-Simulation Model (MSSM) is available at \href{https://sites.google.com/view/yongseok/data-access}{https://sites.google.com/view/yongseok/data-access}. \label{foot:href}}.

\vspace{2mm}
\subsection{Is The Machine-assisted Semi-Simulation Model (MSSM) Compatible With Semi-Analytic Models (SAMs)? }
\label{sec:result-compatible}
 
In Figure  \ref{fig:mdpl_pred}, for six baryonic properties we estimate, we compare the PDFs of our machine learning model ({\it red solid lines}), and of a SAM ({\it green dotted lines}).
For a representative SAM, we utilize the MDPL2-{\sc Sag} catalogue  ({Cora et al. \citeyear{cora2018SAG}}), one of the three SAM-generated galaxy catalogues in the {\sc MultiDark-Galaxies} database ({Knebe et al. \citeyear{knebe2018multidark}}).\footnote{The {\sc MultiDark-Galaxies} data can be found in the {\sc CosmoSim} database at http://www.cosmosim.org/.}      
We also add the {\it actual} baryon data in the {\sc IllustrisTNG} for comparison (TNG100-1; {\it black dashed lines}).  
Overall, we find that our MSSM and the SAM ({\sc Sag}) exhibit largely compatible distribution functions.  
For certain properties like black hole masses, star formation rate, and stellar magnitudes, there is a sign that the MSSM mimics the distribution of {\sc IllustrisTNG} more closely --- which is what MSSM is specifically designed to do.    
Yet, there are some clear mismatches due in large part to the small number statistics.  
For example, in the gas mass distribution, at $M_{\rm gas} \lesssim 10^{9.5}\,\msun$, the MSSM's prediction deviates from {\sc IllustrisTNG} leading to a sizable gap at the lowest mass end ({\it 1st row, left panel}). 
The MSSM's prediction for metallicity drops drastically at $\log({\rm Metallicity}) \gtrsim -1.8$, too ({\it 3rd row, left panel}). 

We then consider the relation between the predicted stellar mass and the halo mass, $M_{\star} - M_{\rm halo}$, in Figure \ref{fig:mstarmhalo}. 
This plot shows how the two halo properties are correlated on a two-dimensional plane (two-dimensional PDF).  
Since stellar mass is one of the properties the machine can estimate well, our MSSM prediction ({\it red-blue contours} in the {\it upper right panel}) replicates the actual $M_{\star} - M_{\rm halo}$ relation in the {\sc IllustrisTNG} run well ({\it top left panel}). 
As a reference, the prediction of three popular SAMs --- {\sc Sag} ({Cora et al. \citeyear{cora2018SAG}}), {\sc Sage} (Croton et al. \citeyear{croton2016SAGE}), and {\sc Galacticus} ({Benson \citeyear{Benson2010-wy}}) --- are shown here as {\it gray contours} demarcating $\rho_{\rm bin,\, cutoff} = 0.01\, /({\rm log}_{10}\,{\rm M}_{\sun})^{2}$ (see Figure 8 of {Knebe et al. \citeyear{knebe2018multidark}}).
Also as a reference, added to Figure \ref{fig:mstarmhalo} is the result of the baseline model ({\it bottom right panel;} see Section \ref{sec:extract} and Table \ref{table:input}). 
Because of various improvements, our MSSM tends to perform better in the lower mass range (say, $M_{\star} < 10^{9.5}\,\text{M}_{\sun}$) than the baseline model does.  

As illustrated in Figures \ref{fig:mdpl_pred} and \ref{fig:mstarmhalo}, we find that the MSSM pipeline can be a promising way to  transplant the baryon physics of a high-resolution galaxy-scale hydrodynamic simulation (e.g., {\sc IllustrisTNG}) onto a larger-volume DM-only simulation (e.g., {\sc MultiDark-Planck}).
It is also worth noting that our machine can ``paint'' galaxies and their baryonic properties on a large $(1 \,\,h^{-1}{\rm Gpc})^3$ DM-only run, within a fraction of time required for a high-resolution hydrodynamic calculation --- a few tens of minutes (at most) versus a few weeks (at least).  

\begin{figure*}
    \includegraphics[width=0.88\textwidth]{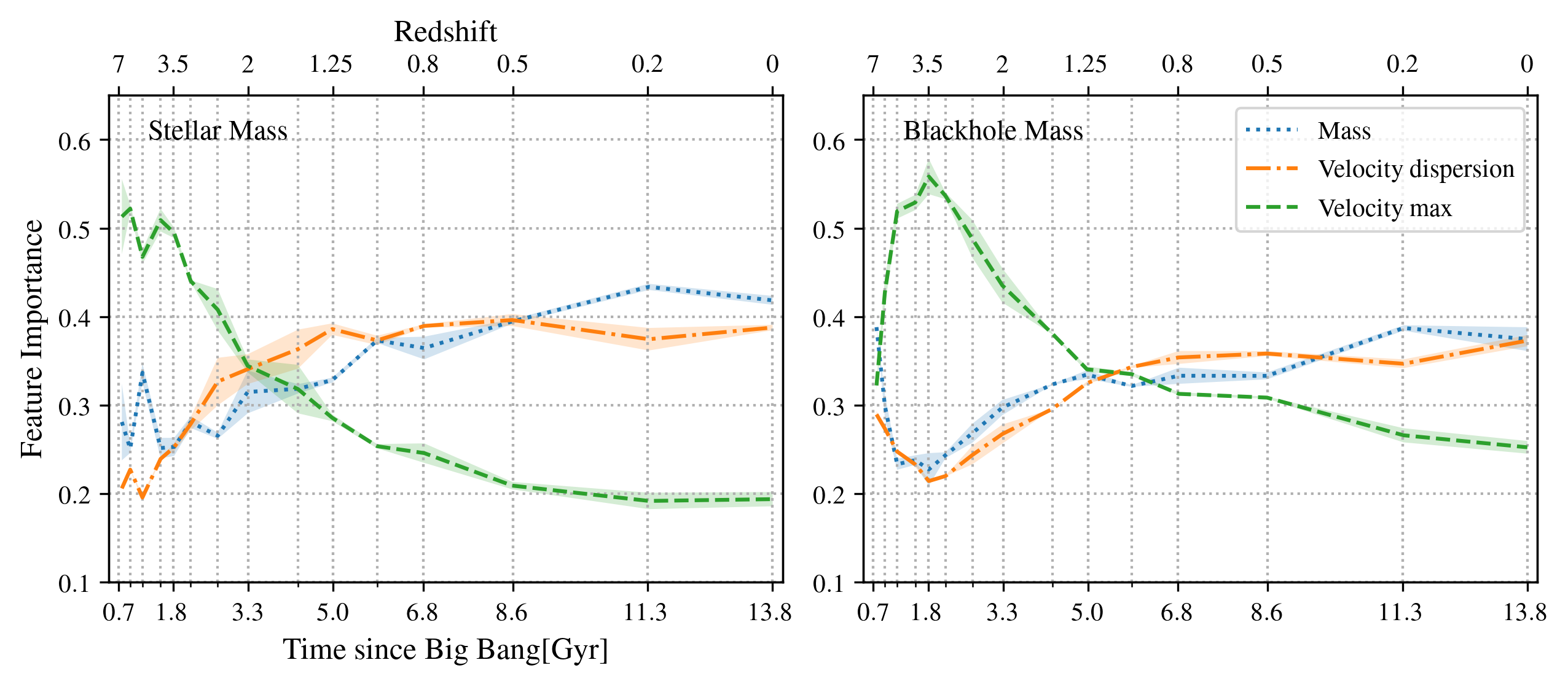}
    \caption[width=\textwidth]{Relative importances of input features --- halo mass, velocity dispersion, maximum circular velocity --- when the machine predicts stellar masses ({\it left panel}) and central black hole masses ({\it right panel}) based only on the three DM features of halos in {\sc IllustrisTNG} (i.e., baseline model; see Section \ref{sec:extract} and Table \ref{table:input}).  
    The evolution of the feature importances are plotted as functions of time.
    See Section \ref{sec:fimp} for more discussion about this figure.     
    }
    \label{fig:fimp}
    \vspace{0mm}    
\end{figure*}

\subsection{Where The MSSM Can Be Improved}
\label{sec:result-underfit}

In Figure \ref{fig:ssfr_starmass}, we plot the probability distribution of halos on the plane of predicted stellar masses and predicted specific star formation rates (sSFR). 
Shown in each panel is the MDPL-{\sc Sag} catalogue ({\it black dotted contours;} the outermost contour marks $\rho_{\rm bin,\, cutoff} = 0.05\,/({\rm log}_{10}\,{\rm M}_{\sun} {\rm log}_{10}\,{\rm yr}^{-1})$) which best matches the observational data ({\it black circles;} Elbaz et al. \citeyear{Elbaz2011-eu}) among SAMs; see Figure 3 of {Knebe et al. \citeyear{knebe2018multidark}}. 
Notice that the {\sc IllustrisTNG} data itself ({\it red-blue contours} in the {\it left panel}) slightly underpredicts the Elbaz et al. \citeyear{Elbaz2011-eu} data at a given stellar mass when compared with MDPL-{\sc Sag}, but better matches the {\it GALEX}-SDSS-{\it WISE} Legacy Catalog ({\it black diamonds}; Salim et al. \citeyear{salim2016}).
The MSSM prediction behaves in a similar way ({\it red-blue contours} in the {\it right panel}), which is again exactly what the MSSM is trained to do.
However, the two-dimensional distribution of halos is narrower in machine predictions than in the original  {\sc IllustrisTNG} data, as is indicated by the smaller error bars for the binned averages ({\it yellow squares} in the {\it right panel}).  
A similar tendency is found in Figure \ref{fig:mstarmhalo} as well, where the halos are distributed in a narrower strip in MSSM predictions but not as much.  
When only one axis is of a predicted property (e.g., Figure \ref{fig:mstarmhalo}), the two-dimensional distribution seems broader than when both $x$- and $y$-axis are of predicted properties (e.g., Figure \ref{fig:ssfr_starmass}).

The narrower distribution of halos likely implies reduced diversity of galaxies of same stellar masses. 
We suspect that when the machine is asked to predict baryonic features from DM-related features only, it may have been underfitted due to the inherently limited number of available input features.  
That is, there are only a very few {\it important} input features that decides the output, so the diversity of resulting outputs is highly restricted (more discussion in Section \ref{sec:fimp}).  
This is the area where our MSSM pipeline should and can be improved in future studies (see Section \ref{sec:future}).  

\vspace{-3mm}

\section{DISCUSSION}
\label{sec:discuss}

In this section, we discuss two topics we find useful to appreciate our MSSM pipeline and its scientific usages.  

\subsection{Relative Importance of Input Features}
\label{sec:fimp}

Since our machine is built with ERT, a RF-type learning algorithm, we can easily find which input feature contributes more than the others (e.g., halo mass vs. halo angular momentum) in estimating a particular halo property (e.g., stellar mass). 
The degree of contribution by each of the input features is termed ``feature importance''.  
Feature importance is a relative metric among all input features adopted, and lie in the range of [0, 1]. 
For example, the feature importances of input parameters $P_1$, $P_2$, $P_3$ could be 0.6, 0.3, 0.1, respectively, which add up to 1. 

Figure \ref{fig:fimp} shows how relative importances of input features in the baseline model (see Section \ref{sec:extract} and Table \ref{table:input}) change over time when predicting  two baryonic properties. 
At high $z$, the maximum circular velocity is the most responsible in constructing the mappings from inputs to outputs --- to both stellar mass ({\it left panel}) and central black hole mass ({\it right panel}). 
However, at lower $z$, the halo mass and velocity dispersion take over and become more dominant.  
The trends robustly appear across 15 redshift snapshots from $z=7$ to 0 we tested, and are highly similar for both mass predictions.  
At $z=0$, the halo mass is the most important feature in estimating both properties with features importances $\gtrsim 0.4$. 

From feature importances we expect to extract physical insights about how cosmological structures have formed and evolved.  
We may also use features importances to evaluate how effective a new input feature is compared to preexisting ones.    
For example, a similar test with our improved MSSM reveals that the three input features shown in  Figure \ref{fig:fimp} are still more important than most other newly introduced features in Table \ref{table:input} (or see Section \ref{sec:extract}) most of the time. 
To raise the scientific potential of MSSM, our next goal would be to develop a set of new inputs whose feature importances  are comparable to the three existing ones'.  

\subsection{Required Training Set Size To Build MSSM}
\label{sec:learning}

\begin{figure}
    \vspace{2mm}
    \includegraphics[width=\linewidth]{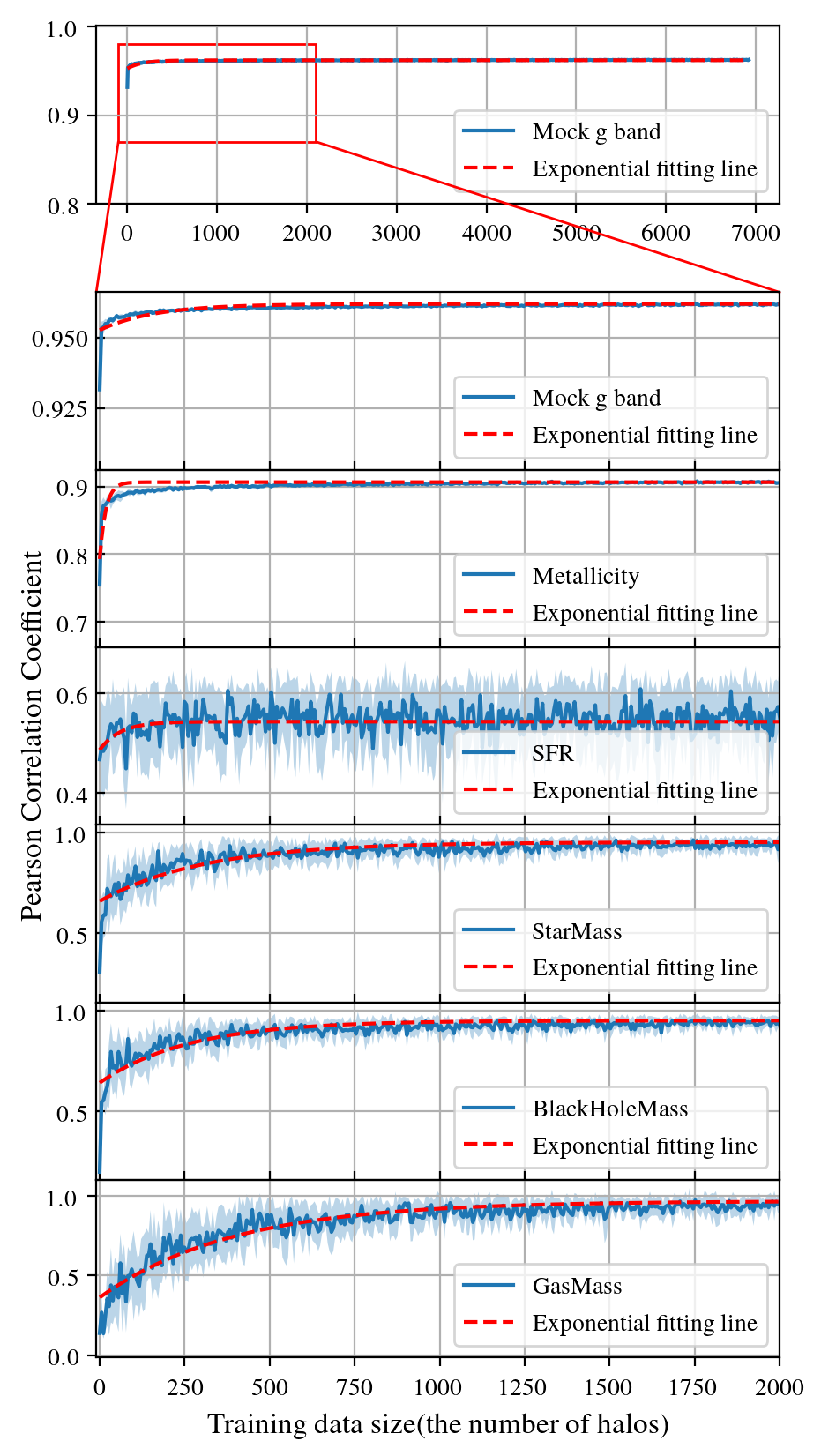}
    \vspace{-2mm}
    \caption[width=\linewidth]{Effect of a training set size on the machine accuracy, Pearson correlation coefficient (PCC), Eq. (\ref{eq:PCC}), when the machine predicts various baryonic properties (each of {\it six panels}) based on three DM features of halos in {\sc IllustrisTNG} (i.e., baseline model; see Section \ref{sec:extract} and Table \ref{table:input}).
    The ``learning curves'' reach their maximum accuracies with only $\lesssim$$10^3$ halos in the training set.      
    See Section \ref{sec:learning} for more discussion about this figure.     
    }
    \label{fig:learning_curve}
    \vspace{-3mm}    
\end{figure}

Generally speaking, the size of a training set is one of the deciding factors in the quality of supervised learning.  
To check whether our TNG100-1 training set (Section \ref{sec:TNG100}) is sufficiently large, we measured the machine accuracy with PCC, Eq. (\ref{eq:PCC}),  as we increase the size of the training set.  
In Figure \ref{fig:learning_curve}, we see the effect of the training set size on the accuracy of the baseline model (that uses just three input features --- halo mass, velocity dispersion, maximum circular velocity; see Section \ref{sec:extract} and Table \ref{table:input}).  
Readers may notice that for all six baryonic properties we estimate, the ``learning curves'' reach their maximum accuracies with only a surprisingly small number of halos in the training set.  
For example, for stellar mass and gas mass predictions, $\sim$$10^3$ halos are enough to yield reasonably good estimates.   
For stellar magnitudes ($g$ band) and metallicities, $\sim$$10^2$ halos seem sufficient for the machine to reach its maximum potential.
From the shapes of learning curves one may argue, for example, that the stellar magnitudes are highly correlated with the three input features (steep ascent to PCC $\sim$ 1 only with $\sim$$10^2$ halos), or that SFR is relatively hard to predict no matter how many halos are used in training (steep ascent but only to PCC $\sim$ 0.5).  

The baseline model can be well-trained up to its full potential with just $\lesssim$$10^3$ halos, at least for the presented machine learning algorithm. 
Because the $z=0$ training set from TNG100-1 even after aggressive data pruning (Section \ref{sec:prune}) still holds $\sim4\times10^4$ halos, the machine trained with TNG100-1 can be considered to have reached its maximum accuracy.\footnote{To doubly ensure that our $z=0$ training set is sufficiently large, we trained a machine with all nine halo catalogues within $z<0.1$.  Using a $\sim$9 times bigger training set did not significantly improve the machine accuracy, as expected by the saturated learning curves in Figure \ref{fig:learning_curve}.}  
We suspect that if the machine is built with more {\it important} input features (i.e., not just three features in the baseline model; see Section \ref{sec:fimp}), a bigger training set would be needed to converge to the maximum accuracies in the learning curves.  
Combined with what we see in Sections \ref{sec:result-underfit} and \ref{sec:fimp}, our experiments suggest that the machine's accuracy is  limited not necessarily by the data size available for training, but more likely by the number of {\it important} input features.  
We will discuss more on potential ways to improve the machine in Section \ref{sec:future}.

\section{CONCLUSION}
\label{sec:conclusion}

\subsection{Summary}
\label{sec:summary}

Using machine learning techniques, we have developed a pipeline to estimate baryonic properties of a galaxy based purely on DM-related features of its host halo (Section \ref{sec:method}).  
Our MSSM pipeline was trained with the {\sc IllustrisTNG} high-resolution hydrodynamic simulation of a $(75 \,\,h^{-1}{\rm Mpc})^3$ volume, so it can establish correlations between DM and baryonic properties (Figure \ref{fig:structure}).  
Compared to a simpler baseline model similar to prior studies, our machine's accuracy has been significantly improved by several improvements --- such as a refined error function with logarithmic scaling in machine training, considering historical and environmental factors of a halo as inputs, and the two-stage learning with stellar magnitudes as an intermediary.  
Machine accuracies by each and combinations of these improvements were extensively discussed (Sections \ref{sec:preprocess} and \ref{sec:result1}).  
For example, the logarithmic scaling in the error function alleviates the inaccuracy in the lower end of the predicted gas masses. 
The two-stage learning in which predicted stellar magnitudes from one machine is used as an input in the next, is found to be very effective in increasing the prediction accuracy for SFRs.  

 

Once a well-trained machine is in place, in just a few tens of minutes we can rapidly populate a DM-only simulation volume that is large enough to address topics like baryonic acoustic oscillations, with galaxies having basic properties.   
With our MSSM mimicking {\sc IllustrisTNG}'s galaxy-halo correlation better than previous models, we painted baryonic properties on DM halos in a $(1 \,\,h^{-1}{\rm Gpc})^3$ volume of the {\sc MultiDark-Planck} DM-only simulation (Section \ref{sec:result2}).   
The resulting MSSM galaxy catalogue\textsuperscript{\ref{foot:href}} is largely compatible with popular SAM catalogues. 
Furthermore, our MSSM has multiple scientific advantages:  

\vspace{-1mm}
\begin{itemize}
\item (1) Within a fraction of time needed for a hydrodynamic simulation, one can efficiently transplant the baryon physics of galaxy-scale hydrodynamic calculations onto a much larger volume.  
Readers should note that, unlike SAMs, this process does not require any recipes with fine-tuned parameters or human bias.  
\item (2) The ERT algorithm naturally assesses the relative importances of input features in estimating each baryonic properties (Section \ref{sec:fimp}).  
The feature importance enables us to select important input features easily, and refine the machine with newly added input features with higher importance scores.
\item (3) From feature importances, and by comparing the MSSM catalogue\textsuperscript{\ref{foot:href}} with SAMs', one can expect to discover physical insights in structure formation and improve the physics models in SAMs. 
\end{itemize}

Despite the many improvements we made over the baseline model, clearly there is room for further improvements for our MSSM framework. 
In the present paper, we have assumed that dark matter properties of {\sc IllustrisTNG} and {\sc MultiDark-Planck} are largely similar that we can ignore the baryonic back reaction. 
But it may introduce inaccuracy in baryon-rich halos (see Appendix \ref{app:baryon} for more discussion). 
Additionally, the lack of diversity discussed in Section \ref{sec:result-underfit} needs to be addressed by, for example, finding new input features that are better correlated with a desired output (see Section \ref{sec:future} for more discussion).

\subsection{Future Work}
\label{sec:future}

A well-constructed machine that finds correlations between DM and baryonic contents could open up a new window to understand how our Universe has evolved.
Despite important progresses we have made, immediate future projects as well as areas of improvements  still remain.  

\vspace{-1mm}
\begin{itemize}
\item (1) The analysis in Section \ref{sec:result2} compares only the scalar properties of galaxies generated by MSSM and SAMs.  
In the subsequent study, we compare the spatial distribution of MSSM galaxies with the {\sc MultiDark-Galaxies} data ({Knebe et al. \citeyear{knebe2018multidark}}).
\item (2) The hyper-parameter space of ERT has not been fully explored.\textsuperscript{\ref{foot:hyper}}
We may need to develop a more sophisticated error function for ERT to capture the diverse nature of correlations  --- not simply linear but complex in a multi-dimensional way --- between inputs and outputs.  
By exploring and tuning the hyper-parameters, we may resolve the underfitting issue described in Section \ref{sec:result-underfit}. 
\item (3) As noted in Sections \ref{sec:result-underfit} and \ref{sec:discuss}, the accuracy of the proposed machine-based approach is likely limited by the small number of {\it important} input features.  
To raise the scientific potential of MSSM, we will need to find new {\it important} input features.  
For example, a set of features characterizing the merging event can be useful --- not just the mass ratio, but e.g.,  collisional orbit parameters, infall rates, etc. 
These new input features will need to be extracted not from the halo catalogue or merger trees, but from a sequence of simulation snapshots finely spaced in time.  
One may apply a convolutional neural network to the simulation sequence itself to learn and predict baryonic properties (somewhat similar to Zhang et al. \citeyear{zhang2019}). 
\end{itemize}

\vspace{-3mm}
\section*{Acknowledgments}

The authors thank Yun-Young Choi, Harshil Kamdar, Juhan Kim, Joel Primack, and the anonymous referee for insightful discussion and feedback on our research.
Ji-hoon Kim acknowledges support by Research Start-up Fund for the new faculty of Seoul National University (SNU), and by Creative-Pioneering Researchers Program through SNU.
This work was also supported by the National Institute of Supercomputing and Network/Korea Institute of Science and Technology Information with supercomputing resources including technical support, grants  KSC-2018-S1-0016 and KSC-2018-CRE-0052. 
The {\sc CosmoSim} database used in this paper is a service by the Leibniz-Institute for Astrophysics Potsdam (AIP).
The {\sc MultiDark} database was developed in cooperation with the Spanish MultiDark Consolider Project CSD2009-00064.

\vspace{3mm}

\newcommand{\newblock}{}
\bibliographystyle{mnras}
\bibliography{refs}

\appendix

\vspace{2mm}
\section{Verifying Stellar Magnitudes As Information Containers}
\label{app:mag}

Stellar magnitudes play an important role  in the two-stage learning (Section \ref{sec:result-photo}).  
As discussed, stellar magnitudes are found to be a good intermediary between e.g., DM halo mass and SFR.  
Typically, star particles in the simulation are convolved with a stellar population synthesis model (e.g., Bruzual \& Charlot \citeyear{bruzual2003}) and a photometric filter to produce mock band stellar magnitudes.   
Therefore, one may argue that additional astrophysical information is provided to the machine as we utilize stellar magnitudes as an intermediary.  

\begin{figure}
    \centering
    \vspace{-2mm}
    \includegraphics[width=1.09\linewidth]{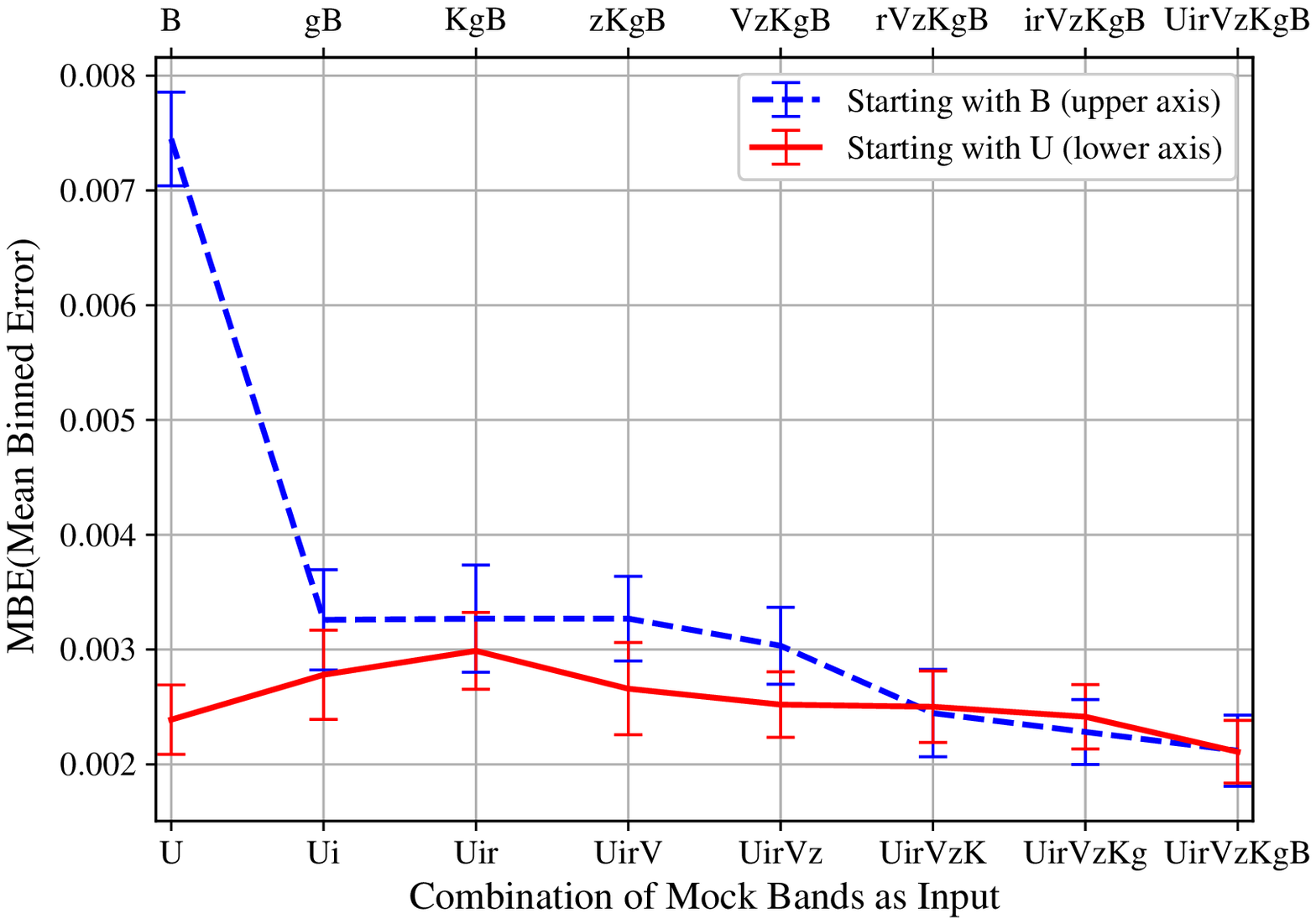}
    \vspace{-2mm}
    \caption[width=\linewidth]{Mean binned error (MBE), Eq. (\ref{eq:MBE}), of stellar mass prediction as a function of how mock band stellar magnitudes are used as an intermediary in the two-stage learning for our MSSM (see Section \ref{sec:result-photo}).\textsuperscript{\ref{foot:trials}}
    Shown on the $x$-axis are various combinations of mock band magnitudes (e.g., ``$zKgB$'' means $z$, $K$, $g$, $B$ bands are used as an intermediary in machine training). 
    The {\it blue dashed line} is for the sequence of combinations shown in the {\it upper axis}, $B$ to $UirVzKgB$. 
    The {\it red solid line} is for the sequence of combinations shown in the {\it lower axis}, $U$ to $UirVzKgB$. 
    This plot demonstrates that $U$ band magnitude is the most dominant feature in predicting stellar mass. 
    The MBE scores are for the entire stellar mass range, not for a smaller range as in Table \ref{table:score}.
    See Appendix \ref{app:mag} for more discussion about this figure. 
    }
    \label{fig:metal_acc}
    \vspace{-2mm}
\end{figure}

To better understand how stellar magnitudes and the two-stage learning help our MSSM to achieve better accuracy, here we evaluate if stellar magnitudes in different bands contain different information.  
In other words, we check if including more photometric bands improves the MSSM's accuracy.  
On the $x$-axis of Figure \ref{fig:metal_acc}, we list combinations of mock band magnitudes used as an intermediary in the two-stage learning when predicting stellar masses. 
For the {\it blue dashed line}, we start with just one band, $B$, and add one more band at a time in the order of $g$, $K$, $z$, $V$, $r$, $i$, $U$ (from left to right on the {\it upper axis}). 
This is the ascending order of feature importance among the eight band magnitudes.   
One can see that as we add more bands, the machine error, MBE, decreases.  
On the other hand, the {\it red solid line} is for the reversed order of combinations starting with $U$ (from left to right on the {\it lower axis}). 
Since the $U$ band magnitude has the highest feature importance, the MBE is already near its minimum only with the $U$ band. 
Adding more bands does not significantly improve the machine accuracy.  

Our tests reveal that for stellar mass predictions the $U$ band is dominant; for metallicity predictions, the $i$ band is. 
Because different band magnitudes carry different information about baryonic physics in a galaxy, we expect that including stellar magnitudes in more photometric bands would improve the MSSM's accuracy.  

\section{Effect of Baryons on Dark Matter Halos}
\label{app:baryon}

\begin{figure}
  \vspace{2mm}
  \includegraphics[width=0.92\linewidth]{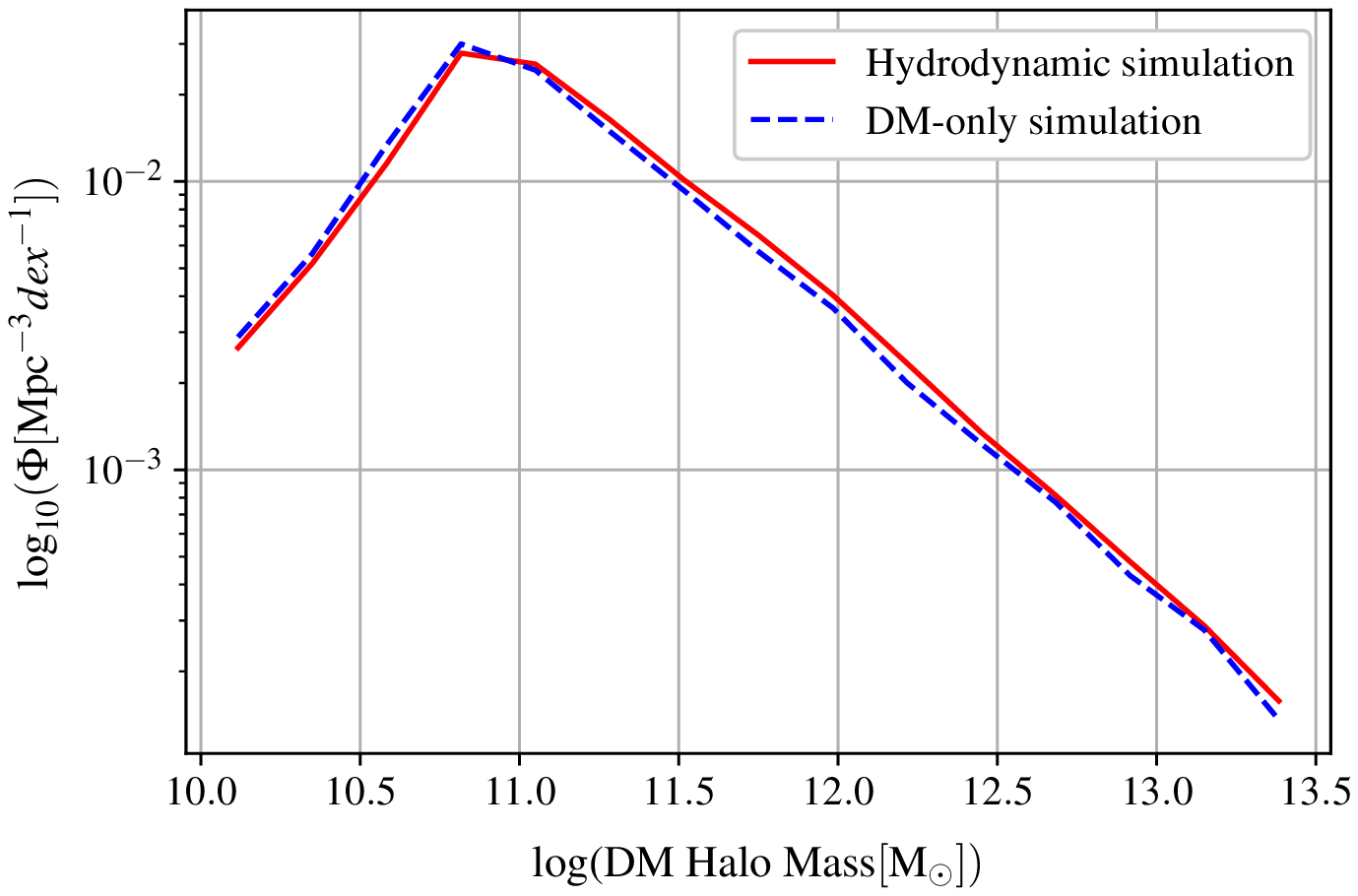}
  \vspace{0mm}
  \caption[width=\textwidth]{ Probability distribution functions $\Phi$ (PDFs) of DM halo masses for {\sc IllustrisTNG} ({\it red solid line}) and {\sc IllustrisTNG-Dark} simulations ({\it blue dotted line}).  The shift between the two lines is only less than 1\%, and can be safely ignored for our purpose when applying our machines.  See Appendix \ref{app:baryon} for more discussion about this figure. 
  }
  \label{fig:dm_compare}
  \vspace{1mm}
\end{figure}

In Section \ref{sec:result2}, we feed a DM-only simulation data to the machine trained with a hydrodynamic simulation data to generate a galaxy catalogue. 
For this to work, an implicit assumption is that DM halos from DM-only simulations and the ones from hydrodynamic simulations starting from an identical IC should have an 1-to-1 match.  
In hydrodynamic simulation, however, the so-called baryon back-reaction may have an effect on the internal properties of a DM halo such as its shape, profile, and circular velocity (e.g., Duffy et al. \citeyear{duffy2010}; Cui et al. \citeyear{cui2012}; Martizzi et al. \citeyear{martizzi2012}; Sawala et al. \citeyear{sawala2013}; Henson et al. \citeyear{henson2017}; Chua et al. \citeyear{chua2019}) and possibly some large-scale properties  (e.g., Cui et al. \citeyear{cui2017}). 
Internal structure of DM halo can also be affected by sophisticated baryonic physics such as AGN feedback. 
In this study, however, we consider only the bulk properties of DM halos such as the ones in Table \ref{table:input}.  
For our MSSM to work, one of the crucial indicators to inspect would be the DM mass function of halos, not the individual internal structures. 
Studies have shown that the DM halo mass function of a hydrodynamic simulation including AGN feedback matches well that of a DM-only simulation (e.g., Duffy et al. \citeyear{duffy2010}; Martizzi et al. \citeyear{martizzi2012}). 
Our own comparison of DM halo mass functions from {\sc IllustrisTNG} and {\sc IllustrisTNG-Dark} (DM-only run of {\sc IllustrisTNG}) in Figure \ref{fig:dm_compare} reveals high resemblance with only a slight shift ($<$1\%). 
For these reasons, we have assumed that DM halos from a DM-only simulation can be used as inputs for a machine trained with a hydrodynamic simulation.  
Further correction and investigation on this issue remains as future work.


\bsp	
\label{lastpage}
\end{document}